\documentclass[useAMS,usenatbib,usegraphicx]{mn2e}

\usepackage{amsmath, hyperref, amssymb, color, graphicx, latexsym}
\usepackage{multirow}
\usepackage{xspace}
\usepackage{algorithm, algorithmic}
\usepackage{subfigure}

\algsetup{linenosize=\scriptsize, linenodelimiter=.}

\newcommand{\ee}[1]{\!\times\!10^{#1}}
\newcommand{\kepler}{{\it Kepler}\xspace}
\newcommand{\roughly}{\mathchar"5218\relax} 
\newcommand{\snr}{S/N\xspace}
\newcommand{\fluxunits}{$e^-/{\rm s}$\xspace} 

\DeclareMathOperator\erf{erf}
\DeclareMathOperator\erfc{erfc}

\title{A Bayesian method for detecting stellar flares}

\author[M.\ Pitkin et al]{
M.~Pitkin,$^1$\thanks{matthew.pitkin@glasgow.ac.uk}
D.~Williams,$^1$\thanks{mail@daniel-williams.co.uk}
L.~Fletcher$^1$\thanks{lyndsay.fletcher@glasgow.ac.uk}
and S.~D.~T.~Grant$^{2,1}\thanks{sgrant19@qub.ac.uk}$ \\
$^1$SUPA, School of Physics and Astronomy, University of Glasgow, University Avenue, Glasgow, G12
8QQ, UK \\
$^2$Astrophysics Research Centre, School of Mathematics and Physics, Queen's University
Belfast, Belfast BT7 1NN, UK}

\date{}

\pagerange{\pageref{firstpage}--\pageref{lastpage}} \pubyear{2014}

\begin{document}

\maketitle

\begin{abstract}
We present a Bayesian-odds-ratio-based algorithm for detecting stellar flares in light curve data.
We assume flares are described by a model in which there is a rapid rise with a half-Gaussian
profile, followed by an exponential decay. Our signal model also contains a polynomial background
model required to fit underlying light curve variations in the data, which could otherwise partially
mimic a flare. We characterize the false alarm probability and efficiency of this method under
the assumption that any unmodelled noise in the data is Gaussian, and compare it with a
simpler thresholding method based on that used in \citet{2011AJ....141...50W}. We find our method
has a significant increase in detection efficiency for low signal-to-noise ratio (\snr) flares. For
a conservative false alarm probability our method can detect 95\% of flares with \snr less than
$\roughly 20$, as compared to \snr of $\roughly 25$ for the simpler method. We also test how
well the assumption of Gaussian noise holds by applying the method to a selection of ``quiet''
\kepler stars. As an example we have applied our method to a selection of stars in \kepler
Quarter 1 data. The method finds 687 flaring stars with a total of 1873 flares after vetos have been
applied. For these flares we have made preliminary characterisations of their durations and
and signal-to-noise ratios.
\end{abstract}

\section{Introduction}\label{sec:intro}

Solar flares have been intensely studied for decades, using imaging, spectroscopy and light-curve
data. The record of soft-X-ray light curves from solar flares, as measured with the Geostationary
Orbiting Environmental Satellites (GOES) extends back in a systematic, well-calibrated record to the
mid-1970s, and from this, and other (shorter) data sets many statistical properties of solar flares
have been deduced, and a fairly solid understanding of the physical processes in solar flares has
emerged. There are other stars, in particular UV Ceti stars, for which large numbers of flares
have been observed and statistics gathered, e.g.\ the observations of \citet{Gershberg:1972} and
\citet{Moffett:1974} used in \citet{Lacy:1976}, or the studies of \citet{Ishida:1991} and
\citet{Dal:2012}. Flares have also been searched for in M Dwarf stars using data from
the {\it Sloan Digital Sky Survey} ({\it SDSS}), using both photometric \citep{Kowalski:2009} and
spectral data \citep{Hilton:2010}. However, the \kepler spacecraft, designed to search for
exoplanets via the transit method, has provided a great new resource of precision long duration
high-temporal resolution data needed to study a large population of flares from a single star, or a
large population of stars. Previous searches for flares in this data set include
\citet{2011AJ....141...50W}, \citet{2012MNRAS.423.3420B}, \citet{2012Natur.485..478M} and
\citet{2013ApJS..209....5S}, which have tended to focus on easily recognized large flares.

This paper describes a method for identifying stellar flares of all sizes, based on their expected
temporal characteristics, using a Bayesian odds ratio method. Our method assumes a specific
flare profile, defined by a Gaussian rise and exponential decay, with a specific range of values,
and is therefore most efficient for flares that can be characterized by this model. However, we also
demonstrate examples of its performance for other flare shapes. The odds ratio provides a
natural quantitative way to discriminate between noise and signal models, which provides additional
information over the by-eye judgement relied upon in previous analyses.

\subsection{White light flares}

Stellar flares are intense, rapid and unpredictable brightenings in the magnetised atmosphere of a
star, caused by the release of previously stored magnetic energy \citep{2010ARA&A..48..241B}. In
the case of solar flares, the usual assumption is that energy is converted to the kinetic energy of
non-thermal particles which stop collisionally and cause heating in the solar atmosphere and
enhanced radiation. Solar flare emission can be detected across the entire electromagnetic spectrum
\citep{2011SSRv..159...19F}. In the case of the Sun the flare excess
spectrum peaks in the optical to near-UV, but optical (`white light') enhancements are hard to
identify in light curves simply because the Sun's quiescent spectrum peaks in the optical as well,
and even the largest solar flares produce an observed optical enhancement which is only slightly
above the background fluctuation caused by solar p-modes \citep{2004GeoRL..3110802W}. When they are
observed, however, solar white light flares provide the most robust means we have of determining the
flare's total radiated energy. Again in the case of the Sun, the optical emission originates from
the lower solar chromosphere and/or photosphere, and tends to have a very impulsive shape, often
showing a rapid rise and fall consistent with fast heating and equally fast cooling, presumably by
radiation in a dense atmosphere \citep{2006SoPh..234...79H}. On the other hand, solar flares
observed at shorter wavelengths,
e.g.\ UV to soft X-rays, have a characteristic fast-rise and exponential decay character, indicative
of rapid heating followed by a combination of conductive and radiative cooling in a more tenuous
coronal plasma. Solar flare white light emission was first observed in 1859
\citep{1859MNRAS..20...13C}, but we still do not have a settled theoretical explanation for its
production. This is partly because, with the difficulty of observing solar white light flares, we do
not have the spectroscopic information required to discriminate between different emission
mechanisms like enhanced blackbody continuum, or free--bound emission. On the other hand, we do
have
exquisite imaging observations to guide us in understanding the structure and development of solar
white light flares, which are of course inaccessible in the stellar case. In the case of stellar
flares we have no such direct spatial information, but spectroscopy of large flares, which dominate
the star's radiative output, can be carried out readily \citep[e.g.][]{2010ApJ...714L..98K}, and
evidence for both enhanced (blackbody) heating and free--bound emission can be found. This type of
study is critical in understanding flare energetics; for example to produce a blackbody or a
free--bound continuum requires heating and ionization in very dense parts of the stellar atmosphere.
This in turn has implications for flare energy transport.

White light flares observed on stars can be at levels very much above the star's quiescent level in
that wavelength, and mostly show the fast-rise exponential decay pattern
\citep[e.g.][]{1976ApJS...31...61M, 2013ApJS..207...15K}. Very large stellar flares are
readily picked out by eye or by relatively simple thresholding methods, but we know from the case of
the Sun that flares occur on all scales of physical size, energy content, and other parameters
\citep{2011SSRv..159..263H}. Identifying the smaller events is very important for understanding
magnetic energy release in general, therefore we want to have a robust means to search also for
smaller events, not just large ones. Statistical studies of `superflares' on G-dwarfs in the \kepler
sample have been carried out by \citet{2012Natur.485..478M} and \cite{2013ApJS..209....5S} using a
method based on identifying signals above some threshold determined from de-trended signals data.
They find a superflare occurrence rate $dN/dE \propto E^{-\alpha}$ with $\alpha \sim 2$ for all
G-dwarfs, with $dN/dE$ being the rate for flares of total energy $E$. As with all such efforts, the
flare distributions show a turn-over at low energy due to the difficulty of detecting small events,
and the location of this turnover is important in many approaches to fitting the spectral index.
The value of the spectral index in turn determines whether large or small events contribute most to
the overall energisation of the stellar corona, with $\alpha \geq 2$ being required for dominance of
small events - the so-called `nanoflare' heating scenario \citep{1991SoPh..133..357H}. In the case
of the Sun, this $\alpha$ value varies between 1.5 and 2.1, depending on the wavelength in which
observations are made, the way that background is subtracted, and other parameters. Also in the case
of the Sun the flare distributions are produced using radiation signatures which embody only a very
small minority of the radiated flare energy - for example the readily-observed soft X-ray which is
correlated with the total flare energy (identified with the energy content of the fast particles)
but with a rather large scatter \citep{2012ApJ...759...71E}. It would be much more satisfactory to
carry out flare statistics using the energetically-significant white-light radiation, but this is
not possible for the Sun. Therefore we are motivated to develop methods which will allow this to be
attempted for stellar flares, which means we must pay attention to the identification of small
events. We must therefore be able to decide in a robust way whether an excursion in the light curve
is flare-like or a noise fluctuation. We will do this on the basis of the shape of the light-curve,
searching for fluctuations consistent with the fast-rise and exponential decay profile observed in
larger flares. In principle this method could be adapted to search for different shape profiles.

\subsection{\kepler light curves}\label{sec:keplerdata}

\kepler stellar light curves are known to contain low-frequency variability both from the intrinsic
fluctuations of the instrument \citep[see e.g.][]{2010ApJ...713L.120J} and stellar variations
\citep{2010ApJ...713L.155B}. We are not interested in these variations, but if not dealt with
carefully they can influence any flare detection algorithm. The data we have chosen to use is the
{\tt PDCSAP\_FLUX} data. This has had the Pre-search Data Conditioning (PDC) module of the \kepler
analysis pipeline applied, which attempts to remove signatures in the
light curves correlated with the spacecraft and detector and also accounts for discontinuities due
to pixel sensitivity drop-outs \citep[the method attempts to not remove a true astrophysical signal,
although as noted in][some true stellar variability, which we are not interested in, may be
removed]{2011AJ....141...20B}. In the example analysis we will present we have used data from
the \kepler data release 21 described in \citet{KeplerRelease21}. For this release there have been
several improvements over the original version of the PDC \citep{2010ApJ...713L..87J}, to remove
unwanted instrumental artefacts. These improvements mean that the PDC data for the majority of stars
are processed using the multi-scale maximum a posteriori (msMAP) approach described in
\citet{2014PASP..126..100S}, which is an extension to the MAP approach described in
\citet{2012PASP..124..985S, 2012PASP..124.1000S}. It should be noted that in
\citet{2011AJ....141...50W} the ``raw'' data, rather than the PDC data,
are used with the detrending of \citet{2011AJ....141...20B} applied. This means that there is
very different data conditioning between that analysis and ours, making direct comparisons
unreliable as is reflected in our results (see e.g.\ Section~\ref{sec:res2}).

Various detrending methods have been developed to remove the low-frequency instrumental variations
in the light curves (e.g.\ in addition to the methods in the PDC in
\citealp{2010ApJ...713L..87J, 2012PASP..124..985S, 2012PASP..124.1000S, 2014PASP..126..100S}, or the
detrending in \citealp{2011AJ....141...20B}, there is the {\it astrophysically robust correction} of
\citealp{2012A&A...539A.137M, 2013arXiv1308.3644R}) whilst trying to retain astrophysical
variability. These variations often reflect the rotation period of the star and can be due to
the presence of spots on the stars' surfaces. We have examined several methods of processing the
data to remove, or diminish the effects of the large sinusoidal variations seen in the light curves,
which would otherwise hamper a flare detection algorithm.  We wanted a method that we could
automatically apply to all light curves without having to tune it for individual stars. We also
wanted a method that would not add any major extra artefacts into the data or remove significant
power from the flares. Methods such as subtracting a running median, image erosion, whitening the
data with an ``average'' spectrum, high-pass filtering the data, or removing a running polynomial
fit \citep[e.g.\ with the filtering algorithm of][]{Savitzky:1964} can all remove the variations,
but at the cost of adding artefacts around flares (and other transient signals) and diminishing the
signal power. Instead of attempting to remove the variations we have taken the approach of including
them in our model of the data, i.e.\ fitting the variations {\it and} the flare model together. This
is discussed in detail below.

We will be looking at \kepler long cadence data in which there is one photometric data point every
29.42\,mins.  As stated in \citet{2011AJ....141...50W} this is not ideal for the detection of
flares, which generally evolve more quickly than this. However, our method can very easily be
applied to short cadence data.

\section{Detection algorithm}

The general outline of our method is to use Bayesian model comparison to create a detection
statistic. The detection statistic is formed by calculating the ratio of the probability that a
light curve contains a flare-like signal described by a known parameterizable model {\it and} any
background variations (the {\it signal} model) to the probability that the light curve contains just
the background variations {\it or} other non-flare-like signals and the background variations (the
{\it noise} model). We will refer to this ratio as the odds ratio. The statistic is then
characterized by running the algorithm with mock data that contains no signals. The distribution of
the statistic yields a threshold value above which we will consider a flare detected for a certain
false alarm probability (FAP). Using this threshold and sets of mock data containing simulated
signals we can also find the detection probability as a function of signal-to-noise ratio (\snr).

The method of flare detection in \citet{2011AJ....141...50W} involved smoothing the data with a
median filter over a 10\,h interval, finding points that crossed a threshold of 4.5 times the data
standard deviation and then counting a flare as three contiguous threshold crossings. The values
for these three tunable parameters were found by comparing an automated algorithm to results from a
by-eye search on a set of training data. Their method did not make use of the flare signal shape
(i.e.\ information from the below threshold portion of the data was ignored) to try and gain
\snr and therefore lacked sensitivity to smaller flares. In our method we attempt to use
information on the signal shape by creating a parametrized flare model with which to
compare to the data. We also attempt to automatically veto impulsive transients that may be
due to instrumental effects by including them in our noise model. However, we note that these
models can also veto temporally unresolved flares.

If flare light curves were un-modelled, or far less well modelled than we allow for here, there are
other methods based on Bayesian model comparison available. \citet{2008CQGra..25k4038S,
2009CQGra..26o5017S} describe a general method for finding un-modelled bursts based on searching
for excess power (as applied to gravitational wave data analysis, but still more widely applicable).
Whilst \citet{1999A&A...345..121H} have a Bayesian method to search for flares in X-ray data
\citep[based on the Bayesian Blocks method of][]{1998ApJ...504..405S} that looks for change points
in the statistics of the data.

\subsection{The flare model}

\begin{figure}
 \includegraphics[width=0.5\textwidth]{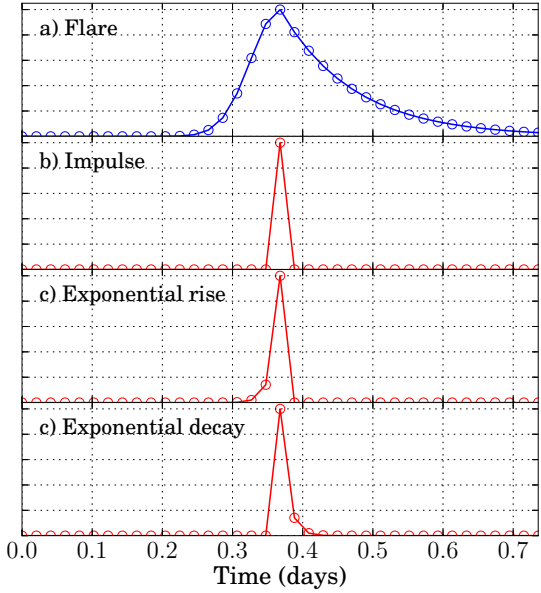}
 \caption{\label{fig:signalmodels} Examples of the signal and noise transient models used in the
algorithm. Panel a) give the flare signal model and panels b), c) and d) give the noise transient
models.}
\end{figure}

The simple flare model we use is based on the observed shape of many flares, with a fast rise
and exponential decay \citep[e.g.][]{2011ASPC..448.1157K}. The rise stage is modelled by a
half-Gaussian, whilst the decay stage is an exponential fall
\begin{equation}\label{eq:flaremodel}
m(t, \tau_g, \tau_e, T_0) = A_0
  \begin{cases}
    e^{-(t-T_0)^2/(2\tau_g^2)} & \text{if } t \le T_0, \\
    e^{-(t-T_0)/\tau_e} & \text{if } t > T_0,
  \end{cases}
\end{equation}
where $A_0$ is the amplitude at the flare peak time of $T_0$, $\tau_g$ is the standard deviation of
the Gaussian rise and $\tau_e$ is the exponential decay time constant. The parametrisation also
allows estimation of these parameters from the detection algorithm. An example of the flare model
is given in Fig.~\ref{fig:signalmodels}. We note that this model does not account for all
potential flare shapes, e.g.\ the ``gradual'' flares described in \citet{2011ASPC..448.1157K}, but
still has some power to detect flares with different morphologies (see Section~\ref{sec:shapes}).

\subsection{The detection statistic}

To create a detection statistic we follow a similar method to that used to search for ring-down
gravitational wave signals developed in \citet{2007PhRvD..76d3003C}. In the simplest terms we want
to test the hypothesis that the data, $d$, contain a flare signal and some background noise (the
form of which we discuss below), $\mathcal{F}$, compared to one in which the data just consists of
the background noise, $\mathcal{N}$. As we will see below this can be extended to include extra
models as required. We can make this comparison by calculating the so-called odds ratio
\begin{equation}\label{eq:oddsratio}
\mathcal{O} = \frac{p(\mathcal{F}|d)}{p(\mathcal{N}|d)} =
\frac{p(d|\mathcal{F})}{p(d|\mathcal{N})}\frac{p(\mathcal{F})}{p(\mathcal{N})},
\end{equation}
where the first fraction on the right hand side is known as the Bayes factor, and the second part is
known as the prior odds, i.e.\ the prior belief in each hypothesis. From Bayes theorem we can
calculate the posterior probability distribution of a set of parameters $\vec{\theta}$ defining the
model in a hypothesis $H$, given a set of data $d$, via
\begin{equation}\label{eq:bayes}
p(\vec{\theta}|d,H) = \frac{p(d|\vec{\theta},H)p(\vec{\theta})}{p(d|H)},
\end{equation}
where $p(d|\vec{\theta},H)$ is the likelihood of the data given $H$ and $\vec{\theta}$,
$p(\vec{\theta})$ is the prior probability on those parameters, and $p(d|H)$ is the probability of
the data given the hypothesis (this is a normalisation factor often known as the Bayesian evidence).
To get to equation~\ref{eq:oddsratio} the evidence value for each hypothesis must be calculated,
which can be performed via marginalisation (which is just another term for integration) of the
product of the likelihood and prior over the model parameters
\begin{equation}
p(d|H) = \int^{\vec{\theta}} p(d|\vec{\theta},H) p(\vec{\theta}) {\rmn d}\vec{\theta}.
\end{equation}
In our detection algorithm we assign the same prior probability to the two (or more) hypotheses and
therefore set the prior odds to be one, so the odds ratio is entirely defined by the evidence values
(i.e.\ it is given by the Bayes factor).

For our algorithm we will assume a Gaussian likelihood distribution for the noise in the light curve
data, which means that for a generic model $m$ parametrized by $\vec{\theta}$ the likelihood of the
data given the model parameters is
\begin{equation}\label{eq:likelihood}
p(d|\vec{\theta}) = \frac{1}{(2\pi\sigma^2)^{n/2}}\exp{\left(-\sum_{j=1}^{n}
\frac{[d(t_j)-m_j(\vec{\theta},t_j)]^2}{2\sigma^2} \right)},
\end{equation}
where $\sigma$ is an estimate of the noise standard deviation (here assumed to be constant over the
data) and $n$ is the number of data points. If we assume that any light curve is purely described by
Gaussian random noise and flare signals (i.e.\ for the moment we ignore any other low-frequency
variability or correlated noise effects) then for $\mathcal{F}$ the model $m$ is given by
equation~\ref{eq:flaremodel} (so $\vec{\theta} = \{A_0,\tau_g,\tau_e,T_0\}$) and for $\mathcal{N}$
we have $m=0$. For our detection statistic we do not care about the actual value of the model
parameters other than the flare time, so we can marginalize over a subset of the parameters
$\vec{\theta}' = \{A_0, \tau_g,\tau_e\}$. However, as discussed in Section~\ref{sec:pe}, we can
still recover $A_0$ when performing parameter estimation. Inserting equation~\ref{eq:likelihood}
into equation~\ref{eq:oddsratio} as appropriate gives an odds ratio as a function of the flare time
of
\begin{align}\label{eq:unmarged}
  \mathcal{O}(T_0) =& \int^{\vec{\theta}'} \exp{}\Bigg( \frac{1}{2\sigma^2} \Bigg[ \Bigg\{
  2\sum_{j=1}^n m_j(\vec{\theta}',T_0)d_j \Bigg\} \nonumber \\
  & - \Bigg\{ \sum_{j=1}^n m_j(\vec{\theta}')^2 \Bigg\} \Bigg] \Bigg) p(\vec{\theta}') {\rm
  d}\vec{\theta}',
\end{align}
where the subscript $j$ refers to the data or model value at the time $t_j$ (note that terms
involving $d^2$ and the pre-factors have cancelled out in forming the ratio of likelihoods).

This assumption of a Gaussian likelihood is mainly due to its simplicity, but generally it
represents the least informative probability distribution for noise \citep[see
e.g.\ chapter 2 of][]{Bretthorst:1988}. Noise in real \kepler light curves is not purely Gaussian
(see Section~\ref{sec:real}), so some care does need to be taken in estimating noise levels and
assessing results when this assumption is used.

Assuming a constant prior probability on the signal amplitude $A_0$ (which we will discuss more in
Section~\ref{sec:others}) the integral in equation~\ref{eq:unmarged} is analytic over $A_0$
between $[0, \infty]$, giving
\begin{align}\label{eq:margA0}
\mathcal{O}(T_0) =& \int_{\tau_g^{\rm min}}^{\tau_g^{\rm max}} \int_{\tau_e^{\rm
min}}^{\tau_e^{\rm max}}
\exp{\left(\frac{D^2}{2\sigma^2 M}\right)}\sqrt{\left(
\frac{\pi\sigma^2}{2M} \right)}\times \nonumber \\
&\left( 1 + \erf{\left[ \frac{D}{\sqrt{2\sigma^2M}}\right]} \right) p(\tau_g, \tau_e)p(A_0) {\rmn
d}\tau_g {\rmn
d}\tau_e,
\end{align}
where $M = \sum_{j=1}^n m_j(\vec{\theta})^2$ (where $m$ is the model from
equation~\ref{eq:flaremodel}, but with $A_0 = 1$, i.e.\ independent of $A_0$) and $D = \sum_{j=1}^n
d_j m_j(\vec{\theta})$. To marginalize over $\tau_g$ and $\tau_e$ we just perform the integration
numerically on a grid over the ranges $[\tau_g^{\rm min}, \tau_g^{\rm max}]$ and $[\tau_e^{\rm min},
\tau_e^{\rm max}]$ using the trapezium rule. The grid intervals we have used are discussed in
Section~\ref{sec:paramspace}. For flares we require that the decay is longer than the rise
time-scale ($\tau_e > \tau_g$), but the prior probability distribution for both is otherwise flat,
so the distribution we use is
\begin{equation}\label{eq:timeprior}
p(\tau_g,\tau_e) = \frac{1}{(\tau_g^{\rm max}-\tau_g^{\rm min})(\tau_e^{\rm max}-\tau_e^{\rm
min}) - \frac{1}{2}(\tau_g^{\rm max}-\tau_e^{\rm min})^2}.
\end{equation}
Note that this prior is correct for the times scale ranges we use in this paper, but the time
scale prior area could be a differently shaped polygon for different ranges.

For each $T_0$ value at which we calculate $\mathcal{O}(T_0)$ the summation in $D$ requires $n$
operations, so using the time of each light curve data point as the $T_0$ values would require $n^2$
calculations. However, calculating $D$ for each $T_0$ is just the cross-correlation of the model and
the data, which via the convolution theorem can be calculated using Fourier
transforms\footnote{Cross-correlation of two time series $f$ and $g$ satisfies $f \star g =
\widehat{(\tilde{f}^{\ast} \times \tilde{g})}$, where $\tilde{x}$ and $\hat{x}$ are the Fourier
transform and inverse Fourier transform of $x$ respectively.}, with of order $n\log{}_2n$
operations. This can offer significant speed-up in calculating the odds ratio.

\subsection{A variable background}\label{sec:varback}

As discussed in Section~\ref{sec:keplerdata} real \kepler light curves contain low-frequency
variations. We have chosen to incorporate these variations into our signal model by modelling them
as a polynomial, giving
\begin{equation}\label{eq:polynomial}
 m(t) = m_f(t) + \sum_{i=0}^{N_p} A_i t^i,
\end{equation}
where $m_f$ is our flare model from equation~\ref{eq:flaremodel}, $N_p$ is the number of polynomial
terms, and $A_i$ are the polynomial coefficients. If we treat each of the polynomial coefficients
and our flare amplitude as independent then we can analytically marginalize over them all (see
Appendix~\ref{app:margamp}) leaving us still with only the flare time-scales ($\tau_g$ and $\tau_e$)
to numerically marginalize out to give us a signal odds ratio, $\mathcal{O}_s$. This method requires
no detrending of the data, or applications of offsets, as the effects of these types of
variation are modelled. With $m(t)$ now being our signal model we want to compare this (i.e.\
by forming an odds ratio) with the evidence that the data contains {\it only} a polynomial
background, which can be calculated by setting $m_f = 0$ in equation~\ref{eq:polynomial}.
Technically what we are calculating in each case is an odds ratio for our particular model
($\mathcal{O}_s$ for a flare {\it and} background variability, or $\mathcal{O}_b$, for just
background variability) versus Gaussian noise, but if we then form a ratio of these the Gaussian
noise case cancels out and we get the odds ratio we require.

Due to the fact that the low-frequency variability can have periods down to less than a day, and
individual \kepler quarters span many tens of days, it is not practical to try and use very high
order polynomials to try and fit all the variability. Instead the analysis can be performed on a
sliding window across the data thus allowing a relatively low order polynomial to fit out the
variability. The flare model has $T_0$ centred in the window. The sliding window length needs to be
chosen such that for the range of flare durations searched for it spans the whole flare whilst
also providing enough background on either side of the flare, so that the polynomial does not try
to fit out any of the flare power. We discuss the value we have chosen for this analysis in
Section~\ref{sec:paramspace}.

As this window slides on to and off of either end of the light curve there will not be as many
data points with which to form the odds ratio. To have odds ratios calculated using consistent
amounts of data (which is important when assessing a detection threshold) we cut off odds ratio
values returned within half the window length of either end of the data (i.e.\ when there is not
full overlap between the window and the data).

\subsection{Other models}\label{sec:others}

We know that in \kepler data some stars contain transit signals from exoplanets
\citep[e.g.][]{2013ApJS..204...24B} and eclipsing binaries \citep{2012AJ....143..123M}. Due to the
above method using a sliding window, and attempting to fit background variations, these transits can
occasionally trigger the odds ratio to favour the flare model (this happens as the window starts to
slide on to, and off, the transit). Transits or eclipses could be parameterized and included
in our background model (the numerator in equation~\ref{eq:finalO}), but since transits in \kepler
data are very well studied we propose just vetoing stars with known transits. In the future it could
be that just short stretches of data known to contain a transit are vetoed, or transits/eclipses
are added to the noise model.

\subsubsection{Short transients}\label{sec:noisetransients}

The light curves can also contain impulsing delta-function-like signals, i.e.\ peaks within a single
(or few) time bin(s). These could be caused by short flares that are not temporally resolved into
several bins due to the long cadence of the data, but they could also be instrumental in origin. Due
to this ambiguity we choose to model any such impulse as part of our noise model. We have three
models for such behaviour (see Fig.~\ref{fig:signalmodels}): i) a transient in a single time bin
with a positive or negative amplitude, ii) a transient with a short exponential decay and a positive
amplitude, and iii) a transient with a short exponential rise and a positive amplitude. This
is not an exhaustive list of all potential instrumental artefacts \citep[see e.g.][for
information on various instrumental effects]{KeplerDH} and others may be considered in the future.
For each of these the unknown amplitude can be analytically marginalized over (along with the
variable background polynomial). For model i) there are no other parameters except the peak time and
for each calculation of the sliding window we marginalize the model over all time bins within the
window. As well as vetoing real transients this can get rid of detection artefacts around loud
flares (\snr of a few tens or above) caused by the effects of having a sliding window. For models
ii) and iii), as for the flare we fix the time to be the centre of the sliding window, but have to
marginalize over the short decay time (see Section~\ref{sec:paramspace}). For these noise models we
have odds ratios of $\mathcal{O}_t$, $\mathcal{O}_{e+}$ and $\mathcal{O}_{e-}$ respectively, which
gives a final detection statistic odds ratio of
\begin{equation}\label{eq:finalO}
 \mathcal{O} = \frac{\mathcal{O}_s}{\mathcal{O}_b + \mathcal{O}_t + \mathcal{O}_{e+} +
\mathcal{O}_{e-}}.
\end{equation}

This method biases us against short duration flares, but without some other (instrumental)
information that would allow us to veto transient artefacts this will remain a problem for long
cadence data.

Note also that this algorithm assumes that the data contain just one flare within the sliding
window, whereas in reality it could contain several. This could give a slight bias to the results if
there are close flares as it will reduce the noise model evidence.

\begin{figure*}
\centering
 \subfigure[][]{
 \label{subfig:sim1}
 \includegraphics[width=\textwidth]{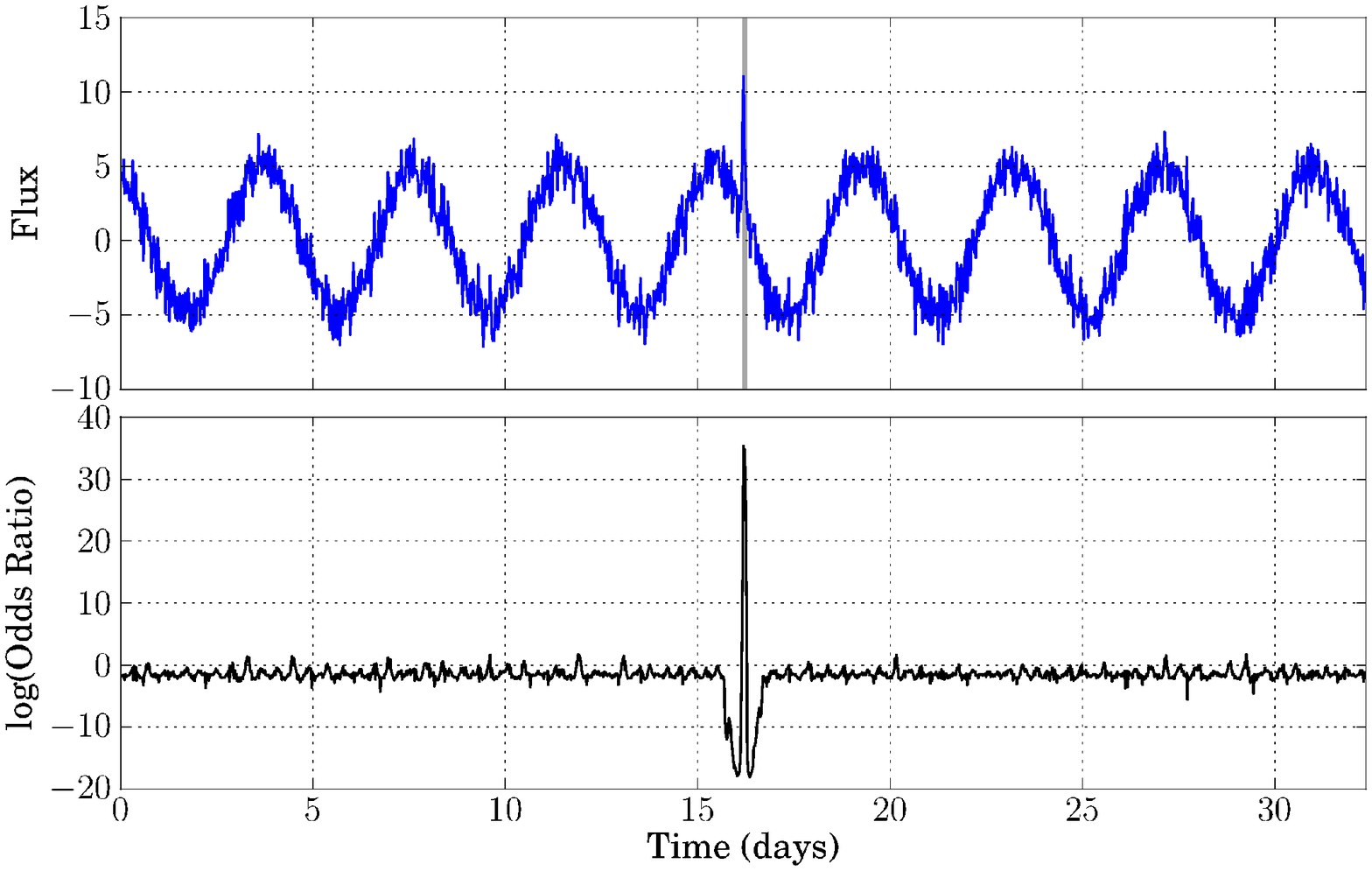}
 }
 \subfigure[][]{
 \label{subfig:sim2}
 \includegraphics[width=\textwidth]{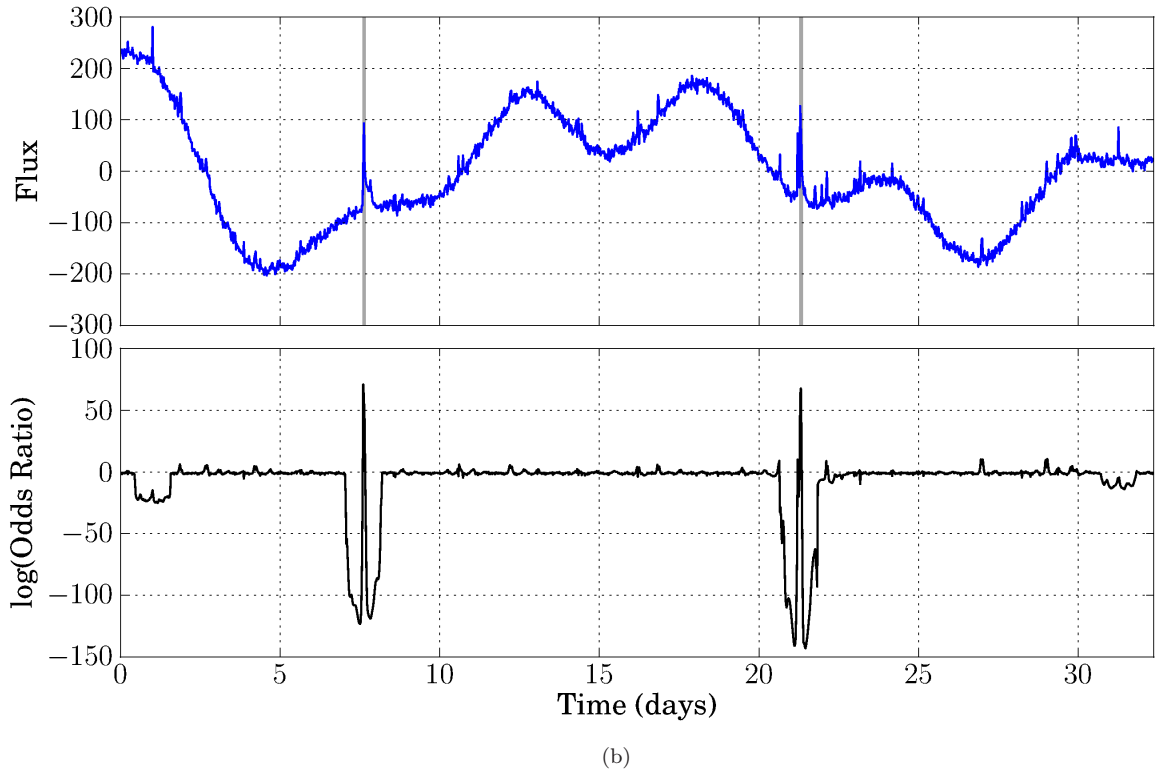}
 }
\caption{\label{fig:examples} Examples of the output of the algorithm using \subref{subfig:sim1} a
simulated light curve containing an injected flare signal, and \subref{subfig:sim2} \kepler Q1 data
for the star with KID 1873543. Grey vertical bars on the light curve plots represent times
when the log odds ratio is above the threshold value of 16.5 (see Sections~\ref{sec:threshold} and
\ref{sec:efficiency}).}
\end{figure*}

Examples of the output of this method, which gives a time series (representing the flare peak time)
of the natural logarithm of the odds ratio (equation~\ref{eq:finalO}), are shown in
Fig.~\ref{fig:examples}. Fig.~\ref{subfig:sim1} shows an example of `mock data' where Gaussian
noise with a mean of zero and standard deviation of unity (in \fluxunits) has been generated, a
sinusoid with frequency 3.9\,d$^{-1}$ has been added along with a flare with parameters
$A_0=10$\,\fluxunits, $\tau_g = 1760$\,s and $\tau_e = 3768$\,s. The algorithm estimated the noise
standard deviation (see Section~\ref{sec:noise}) to be 0.98\,\fluxunits. Figure~\ref{subfig:sim2}
shows the output for \kepler Q1 data for the star with \kepler ID (KID) 1873543. It can be seen that
two flares are obviously found above the threshold set in Sections~\ref{sec:threshold} and
\ref{sec:efficiency}. Conversely, two short duration events seen near the start and end of the light
curve show dips in the log odds ratio. This means they are far more consistent with the noise
models, in particular the short transient noise models. The noise estimate for this data is
8.0\,\fluxunits.

Claiming detection of these signals requires some thresholding on these time series', which we
discuss in Sections~\ref{sec:threshold} and \ref{sec:efficiency}.

\subsection{The parameter space}\label{sec:paramspace}

\begin{figure}
 \includegraphics[width=0.5\textwidth]{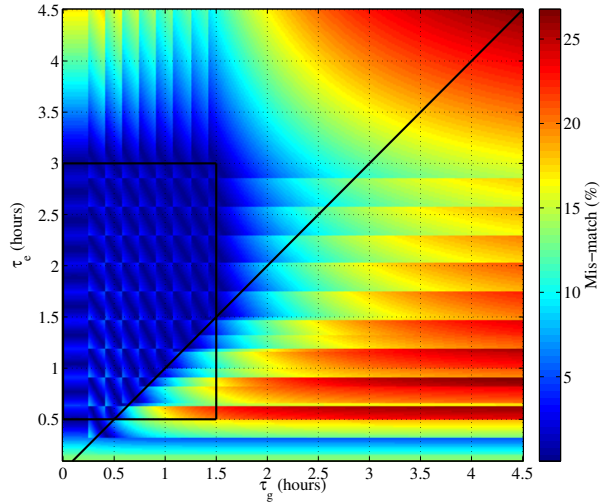}
 \caption{\label{fig:mismatch} The mis-match across the range of flare parameters given the
grid-spacing defined in Section~\ref{sec:paramspace}}
\end{figure}

The ranges of the flare parameters $\tau_g$ and $\tau_e$ that we have chosen cover flares lasting
up to approximately half a day. The Gaussian rise, $\tau_g$, spans from [0, 1.5]\,h, whilst the
exponential decay, $\tau_e$, spans from [0.5, 3]\,h. These ranges go into calculating the
prior given in equation~\ref{eq:timeprior}. Both these ranges are gridded into 10 evenly spaced
points for evaluating and marginalising the odds ratio. If we define the mis-match as being the
fractional power that would be lost by two different flare models (at [$\tau_g$, $\tau_e$]
and [$\tau_g+\Delta\tau_g$, $\tau_e+\Delta\tau_e$] respectively) not completely overlapping as
\begin{equation}\label{eq:mismatch}
M = \left|1 - \frac{\sum_{j=1}^n m_j(\tau_g, \tau_e)m_j(\tau_g+\Delta\tau_g,
\tau_e+\Delta\tau_e)}{\sum_{j=1}^n m_j(\tau_g, \tau_e)^2}\right|
\end{equation}
then we can see how well this gridding covers our parameter space. The mis-match can be seen in
Fig.~\ref{fig:mismatch}, where the black box represents our parameter range, and the diagonal line
gives $\tau_g=\tau_e$. Given that we require $\tau_g < \tau_e$ only values above the diagonal line
are included when calculating our detection threshold. Within our range we find that such a grid
spacing gives a maximum mis-match of less than 10\%, with a mean mis-match of only 1.5\% (i.e.\ a
real flare, that has duration parameters somewhere between our grid points, will be detected with on
average 1.5\% less power than it really has). This level of mis-match assumes that the flares are
fully described by our model, but in reality they are likely to show deviations from this that mean
that no flare will be perfectly matched to our model. Therefore this mis-match is a best case
scenario and more power will be lost for a real search. The grid spacing could be decreased, but
with a corresponding linear increase in computational time of the algorithm.

It can also be seen that even for flare values well outside our range we could still detect them
without losing too much power.

\subsubsection{Amplitude priors}

For the flare amplitudes that we are analytically marginalising over we have assigned a
conservative prior range of between 0 and 1\,000\,000, giving $\log{p(A)} = \log{(10^{-6})} =
-13.8$. As we will assess a detection threshold empirically (see Section~\ref{sec:threshold}) this
prior just provides an overall offset in the odds ratio, so is not of great importance, although we
include it for completeness. The prior ranges for the polynomial amplitudes will cancel between the
denominator and numerator of the odds ratio, so are not required.

\subsubsection{Window length}

As discussed in Section~\ref{sec:varback} the method uses a sliding window. We have chosen a window
$\roughly 27$\,h long (i.e.\ 55 \kepler light curve time bins). This length means that our
longest flares will still have almost all their power within the window, whilst being short enough
that a fourth order polynomial provides a reasonable background fit for variations down to
periods of around two days (see Section~\ref{sec:threshold}).

\subsection{Calculating the noise}\label{sec:noise}

The odds ratio calculation (e.g.\ equation~\ref{eq:margA0}) requires an estimate of the noise
standard deviation $\sigma$ of the data. Just using a standard calculation of $\sigma$ in data
containing flares and other variations, will lead to overestimates of the noise. We instead use a
method to calculate the standard deviation that attempts to veto the effects of flares and the
low-frequency background variations.

The method assumes that the underlying data distribution is Gaussian, with outliers (e.g.\ flares)
in the wings of the distribution. However, as we see in Section~\ref{sec:real}, there are
further correlations in the noise that make this an approximation when using \kepler data. The
low frequency variations in the light curves must initially be filtered out. So, to account for this
we first apply a low-pass filter using the Savitzky-Golay
method mentioned in Section~\ref{sec:keplerdata} with a window and polynomial order the same as used
for our detection algorithm. After this the cumulative probability distribution of the
data is calculated. The standard deviation can then be calculated by finding the values that bound
a certain fraction of the probability distribution around the 50\% value. For example, one standard
deviation ($1\sigma$) should enclose $100\times\erf{(1/\sqrt{2})} = 68.3\%$ of the probability
distribution, so one would find the values $x_{\rm min}$ and $x_{\rm max}$ that bound the
$50-(68.3/2) = 15.85\%$ and $50+(68.3/2) = 85.15\%$ probabilities, and calculate the standard
deviation as $\sigma = (x_{\rm max} - x_{\rm min})/2$. For smaller data sets (and provided the
outliers do not make up a reasonably large fraction of the data) a more accurate value of the
standard deviation can be found by using the value enclosing more of the probability [e.g.\ the
amount enclosing the probability for $2\sigma$, which is 95.4\%, and therefore $\sigma = (x_{\rm
max} - x_{\rm min})/(2\times2)$].

This method is quantitatively similar to repeatedly removing outliers \citep[found with e.g.\
the generalized extreme Studentised deviate test of][]{Rosner:1983} and recalculating the standard
deviation, but does not require repeated iterations. Also, unlike simply removing a certain
fraction of the data in descending order from the largest absolute value first, this method is less
prone towards underestimating the standard deviation when used on data with few, or no, outliers.

This method can potentially lead to an underestimation of the noise if the distribution is not
Gaussian and has broad wings. In \kepler data this could be a problem in cases where there is an
excess of high-frequency correlations. In practice any such light curves could be vetoed during
visual inspection of flare candidate stars. An alternative method, using the power spectrum of the
light curve, and assuming the noise is white and Gaussian at high frequencies, is discussed in
Appendix~\ref{app:noiseest}. However, provided flare signals rise and decay very quickly, this
method is less prone to overestimating the noise than the method described in
Appendix~\ref{app:noiseest} due to power leakage to high frequencies from loud flares.

\section{Algorithm characterization}

Here we will detail the characterization of the algorithm. We will first set a threshold value for
the natural logarithm of the odds ratio (when calculating the odds ratio it is always easier to work
in log-space, and since the logarithm is a monotonic function it makes no difference in practice
using this value) for which we will return a flare detection candidate for a given FAP. Secondly,
we will use this threshold to determine the efficiency of the algorithm for a set of simulated
flare signals. We will also look at how the algorithm performs on real \kepler data to test
the validity of our Gaussian noise assumptions. Other tests are performed to see the effects of
different flare morphologies on our detection ability.

\subsection{Threshold calculation}\label{sec:threshold}

The odds ratio provides a value that, given a set of data, describes the relative probabilities of
competing hypotheses or models. So, provided the prior probabilities for the model parameters and
hypotheses are well defined, an odds ratio of greater than or less than one favours the hypothesis
on the numerator and denominator, respectively. The number of model parameters and their prior
ranges give rise to an Occam factor in the odds ratio calculation (i.e.\ the model with the smaller
parameter space will a priori be favoured due to its simplicity). However, the presence of noise
means that the odds ratio will fluctuate and at low \snr neither hypothesis will be greatly
favoured. \citet{Jeffreys:1931} (in his appendix~B) gives a qualitative assessment of how to
interpret values of the odds ratio, but the significance can also be assessed empirically.

\begin{figure}
 \includegraphics[width=0.5\textwidth]{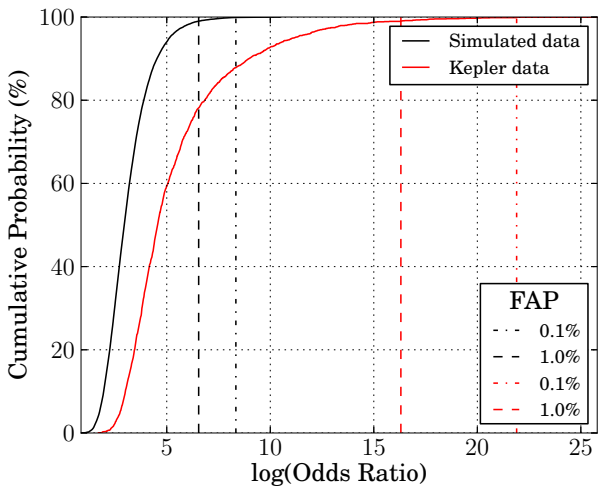}
 \caption{\label{fig:threshold} The cumulative distributions of the log odds ratio for
simulated light curves containing Gaussian white noise and a sinusoidal variation, and real \kepler
data for a selection of ``quiet'' stars. Thresholds for two false alarm probabilities are shown.}
\end{figure}

For a detection algorithm we want to find the distribution of the value of the odds ratio when
looking at data that contain no flares. We can then use this to set a threshold at which we expect
noise alone to exceed this value with a certain probability, known as the FAP. To try and simulate
real data, and incorporate the effect of low-frequency variations, our simulated data sets all
contain sinusoids with amplitudes drawn randomly from a uniform distribution between 10
and 100 times the underlying Gaussian noise standard deviation, with frequencies drawn from a
uniform distribution between 0.5\,d$^{-1}$ and $0.03$\,d$^{-1}$, and an initial phase between 0
and $2\pi$\,rad. We purposely chose this frequency range to be close to the upper end of expected
variations \citep[it is in the tail of the period distribution for solar and late-type stars, e.g.\
Figure 5 of][]{2014ApJS..211...24M}, as from our studies we know that these higher frequency
variations are more likely to produce outliers in the odds ratio distribution (due to the polynomial
background model not fitting the faster changing variations quite so well). This will therefore give
us a reasonably conservative threshold in general. Our simulated data set are based on \kepler
quarter 1 data and are therefore 33.5\,d long, with one point every 29.42\,min. We have run
20\,000 such simulations to derive a threshold. The cumulative probability distribution of the
maximum log odds ratio from each simulation can be seen in Fig.~\ref{fig:threshold}.

If we choose the FAP to be such that there is a 0.1\% chance of noise (Gaussian white noise plus a
background sinusoid), giving one false positive per data set (i.e.\ one
false alarm per thousand light curves, or about one per 80.5\,yr of observations), it gives a log
odds ratio threshold of $8.3$. Equivalent thresholds for false alarm probabilities of 0.2\%, 0.5\%
and 1\% are $7.9$, $7.3$ and $6.5$, respectively. When calculating the efficiency below we calculate
the observed false alarm rate and compare it to these values.

\subsection{Detection efficiency}\label{sec:efficiency}

\begin{figure*}
 \includegraphics[width=1.0\textwidth]{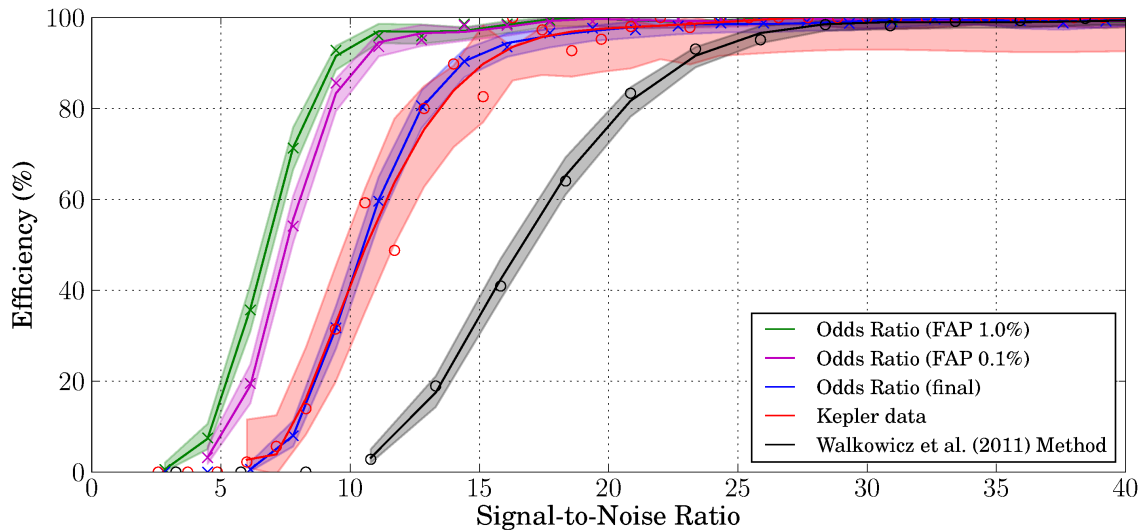}
 \caption{\label{fig:efficiency} The efficiency of our odds ratio algorithm for various
thresholds and the efficiency of our version of the algorithm from \citet{2011AJ....141...50W}. The
efficiency of our algoithm is assessed using simulations in which the noise is Gaussian for
thresholds at 1\% and 0.1\% false alarm probabilities, and at a final threshold of 16.5. The
efficiency based on adding signals to real \kepler data for ``quiet'' stars for a threshold of 16.5
is also shown. The shaded error regions are 95\% confidence intervals around a best-fitting
polynomial, calculated using the Beta distribution described in \citet{2011PASA...28..128C}.}
\end{figure*}

We calculate the efficiency of the algorithm by creating a set of fake signals with varying
parameters and \snr in simulated noise (containing sinusoidal variations as above) and determining
the fraction that are detected above a given threshold. We define a detected signal as a threshold
crossing point (or set of points) that is within two time bins of the known injected central time.
Contiguous above-threshold values are counted as a single detection and any two above-threshold
segments separated by only one time bin are merged into a single detection. We define the
\snr as $\rho = \sqrt{\sum_{j=1}^n m_j^2/\sigma^2}$, where $m_j$ is the injected
signal calculated from equation~\ref{eq:flaremodel} and $\sigma$ is the noise standard deviation
calculated as described in Section~\ref{sec:noise}. We have performed 10\,000 such injections with
\snr spanning between 2 and 50, and $\tau_g$ and $\tau_e$ uniformly drawn from ranges [0,
1.5]\,h and [0.5, 3]\,h, respectively under the condition that $\tau_e \geq \tau_g$ (which is
the same range used by the algorithm described in Section~\ref{sec:paramspace}). The detection
efficiencies for various FAP thresholds are shown in Fig.~\ref{fig:efficiency}. For the threshold
corresponding to a FAP of $0.1\%$ we find detection efficiencies of 50, 95 and 99\% corresponding
to \snr of 7.4, 12.9 and 26.2, respectively. The efficiencies for these, and other, false alarm
probabilities are summarized in Table~\ref{tab:efficiencies}.

\begin{table}
\caption{Efficiency of the detection method for different false alarm
probabilities.}\label{tab:efficiencies}
 \begin{tabular}{@{}lrccc}
\hline
~ & ~ & \multicolumn{3}{c}{\snr for detection efficiency} \\
\hline
~ & ~ & 50\% & 95\% & 99\% \\
\hline
\multirow{3}{*}{FAP} & $1.0\%$ & 6.6 & 10.6 & 25.8 \\
& $0.5\%$ & 7.0 & 11.0 & 26.0 \\
& $0.1\%$ & 7.4 & 12.9 & 26.2 \\
& final & 10.6 & 19.8 & 28.2 \\
\hline
 ~ & ~ & \multicolumn{3}{c}{$\sigma$-threshold method} \\
\hline
 \multicolumn{2}{c}{$4.5\sigma$ threshold} & 16.7 & 25.1 & 34.0 \\
\end{tabular}
\end{table}

When estimating the efficiency we also count any false positives, which are times that the algorithm
exceeded threshold, but for a time that is further than two time bins from the known injected
central time. Given a 0.1\% false alarm threshold of 8.3 and 10\,000 simulations we expect noise
to produce $\roughly 10$ false alarms. However, we find a total of 352 false alarms. Investigating
these we find that 343 of the false alarms are within half of the length of the algorithm's running
window of the injected flare, i.e.\ they are artefacts related to the injections, which if removed
gives a number very close to the expected false alarm rate. Similarly, when calculating efficiencies
for thresholds for false alarm rates of 0.2\%, 0.5\% and 1\%, we actually get numbers of false
alarms of 397, 490, and 669 which, corrected for artefacts, revert to 12, 28 and 85 out of 10\,000
respectively (these suggest that our false alarm probabilities might be slight overestimates).

Given that the algorithm will produce these false positives for loud enough signals it is useful to
determine an odds ratio threshold at which they become significant. Using the distribution of log
odds ratios for the false positives related to flares we can set a new FAP threshold. As these
false alarms are due to flares themselves they will not harm our chances of detection, but will
just slightly bias our number of detected signals. Due to this we are less conservative in setting
the FAP and allow a 1\% value for these artefacts. We find this gives a new threshold of 16.5, for
which the efficiency can also be seen in Fig.~\ref{fig:efficiency}. This gives efficiencies of
50\%, 95\% and 99\% for \snr of 10.6, 19.8 and 28.2, respectively. This is rather conservative
(when just assuming the noise is Gaussian) as it is a FAP that really only applies to
flaring stars rather than all stars, such that for every 100 flares found one is likely to be a
false positive. However, it will also help provide a stronger veto against disturbances in real
\kepler data as shown below. This could be relaxed in the future, especially if further
algorithm development provides stronger artefact vetoes.

\subsection{Characterization using \kepler data}\label{sec:real}

To test the robustness of our assumption that the noise in \kepler light curves is Gaussian we
have run the algorithm on a selection of 2\,000 ``quiet'' \kepler stars with little variability. To
chose this set of ``quiet'' stars we have randomly run through all the Q1 \kepler long cadence light
curves and picked the first 2\,000 fulfilling the following criteria (after the removal of the best
fit quadratic from each light curve): a) the standard deviation of the data before and after
removing a running Savitzky-Golay filter (with a fourth order polynomial and 55 samples in the
running window) are within 25\% of each other, b) the maximum and minimum values within the light
curve are within 7.5 standard deviations of each other (using the standard deviation without the
Savitzky-Golay filtering), and c) the largest peak in the light curves amplitude spectrum (the
square root of the power spectrum) is no more that 7.5 times the median value. These criteria seem
to be successful in selecting stars with very little long duration variability.

For each of these quiet stars we have run our detection algorithm and calculated the maximum
log odds ratio. The cumulative distribution of log odds ratios can be seen
alongside that for purely Gaussian noise (with simulated sinusiodal variations) in
Fig.~\ref{fig:threshold}. It can be seen that real \kepler data have a broader distribution
and are shifted to the right, i.e.\ \kepler data containing no flares more often favour the flare
model than data containing Gaussian noise. We see that odds ratio thresholds for 1\% and 0.1\% false
alarm rates per light curve are $\roughly 16.3$ and $\roughly 21.9$, respectively. The 1\% false
alarm rate threshold is very close to the conservative 16.5 threshold we set in
Section~\ref{sec:efficiency} based on signals themselves producing a 1\% false alarm rate.

To check that these ``quiet'' stars do not actually contain flares we have visually inspected
the light curves for the 17 stars that fall above our previously discussed odds ratio threshold of
16.5. We find that none of these shows signals that conclusively look like flare, but we do see that
the noise can have short-term correlations that can mimic flares.

We have also performed an efficiency test equivalent to the one performed using Gaussian
noise, but with 2\,000 simulated injections added to the ``quiet'' star light curves, and using an
odds ratio threshold of 16.5. This efficiency can be seen in Fig.~\ref{fig:efficiency}. We find
that the efficiency when using real \kepler data is consistent with that when using Gaussian noise,
i.e.\ once you have set a threshold you can recover flare equally well from simulated or real
data, albeit with a potentially different false alarm rate. From these simulations we found 33 false
alarms (or a 1.6\% false alarm rate). This is about the number to be expected given that from
Fig.~\ref{fig:threshold} we see that 16.5 is roughly the 1\% false alarm threshold for \kepler
data, and is also the 1\% false alarm threshold for artefacts associated with the simulated
signals.

Based on this for the results presented in Section~\ref{sec:res} we will use the assumption of
Gaussian noise and a threshold of 16.5, which as shown should lead to a detection false alarm rate
of $\roughly 1\%$.

In the future there are various options for the analysis to try and account for the
non-Gaussianity of \kepler, or other, data. The simplest is to keep the assumption of the data
consisting of Gaussian noise, but explicitly using a detection threshold that is based on real
analysis of ``quiet'' \kepler stars. Another option is to again keep the assumption of Gaussian
noise, but make the method of estimating the noise in the data (see Section~\ref{sec:noise}) give a
more conservative value for its standard deviation. Finally, a method could be developed to model
the correlations in the noise and use this information in the analysis. If the quiet stars truly
are representative of the noise in the majority of stars then they could provide a model of the
noise in the frequency domain that could be used to ``whiten'' all light curves in the analysis.
This model could simply be just an average of all the quiet star spectra, or based on a set of
principal components \citep[e.g.][]{Shlens:2014} calculated from the quiet star noise that capture
the majority of the correlations. We expect the simplicity of the first option (provided quiet, and
non-flaring, stars are easy to identify using the criteria we set out above) or second option makes
them the most viable options in the short term, although future investigations of the third option
may prove illuminating.

\subsection{Method comparison}\label{sec:comp}

It is useful to compare our method with that used in \citet{2011AJ....141...50W}, which defines a
detection as three contiguous positive threshold crossings, where playground data has been used to
set a threshold of 4.5 times the data's standard deviation. We will call this the $\sigma$-threshold
method. In \citet{2011AJ....141...50W} no empirical FAP is given, but if one assumes that the noise
is purely white and Gaussian (which we have shown above is not really the case), then the
expected FAP for a single data set of length $N$ would be given by
\begin{align}
 f &= 2\frac{N}{\sqrt{\pi}}\sum_{i=3}^N \prod_{j=1}^i \int_{\sigma_T}^{\infty} e^{-x^2/2} {\rm d}x,
\nonumber \\
 &= N\sum_{i=3}^N \left(\erfc{\left(\sigma_T/\sqrt{2}\right)}\right)^i,
\end{align}
where $\sigma_T$ is the threshold in number of standard deviations. For $\sigma_T = 4.5$ and
$N=1638$ (the number of points in a \kepler Quarter 1 light curve) this gives a FAP of
$3.1\ee{-14}\%$, i.e.\ for any reasonable data set Gaussian noise should never give a
false detection. If the threshold is dropped to $\sigma_T = 3$ then the FAP becomes $\roughly
2\ee{-6}\%$, which is still very low and far below the level we have set with our algorithm.
However, dealing with real data may well produce a far higher false alarm rate than that given by
these theoretical calculations. This is suggested by the fact that in \citet{2011AJ....141...50W}
the $\roughly 23\,000$ stars searched give 5784 stars with candidate flares, of which visual
checking confirmed 373 with obvious flares and 565 marginal cases. This is a FAP of $\roughly
20\%$. The main thing this again confirms is that \kepler noise is far from ideally Gaussian
as it contains many noise artefacts, and therefore in reality the noise estimate used was likely to
be an underestimate. Our algorithm attempts to account for some of these non-Gaussian noise
artefacts by including them in the noise model, and by using the more robust underlying noise
estimation method described in Section~\ref{sec:noise}.

We have compared the detection efficiency of our method to a version of that used in
\citet{2011AJ....141...50W}, by again simulating 10\,000 data sets containing flares with \snr
between 2 and 100 and time-scale parameters defined in Section~\ref{sec:efficiency}. For this
comparison we have not included any sinusoidal variations in the simulated data sets as we find that
for short period variations the running median smoothing leaves many artefacts in the data (at the
peaks of the variations) that are picked up as false alarms. This may be one of the reasons for the
large false alarm rate seen in the real \citet{2011AJ....141...50W} analysis. As the data are
simulated we have not had to apply any \kepler-like preprocessing, so do not completely represent
the pipeline on \citet{2011AJ....141...50W}. We also estimate the $\sigma$ in the same way as
described in Section~\ref{sec:noise} as we do not know how this was estimated for the original
routine (although it is mentioned that extreme outliers are excluded). As in
\citet{2011AJ....141...50W}, before applying the detection algorithm we subtract a running median
value from the data calculated for a 10\,h window. Fig.~\ref{fig:efficiency} shows detection
efficiencies of 50\%, 95\% and 99\% for \snr of 16.7, 25.1 and 34.0, respectively. It can be seen
that our algorithm substantially improves efficiency over this method, albeit at far higher {\it
theoretical} false alarm rates.

We note that many of the light curve figures shown in \citet{2011AJ....141...50W} seem to have
significantly reduced flare heights than seen in the publicly available data (either the raw
simple aperture photometry data, or the {\tt PDCSAP\_FLUX} data). An example is their fig.\ 2(a)
(KID 10320656) for which the flare at $\roughly 5$\,d is reduced by an order of magnitude and
an obvious flare at $\roughly 25$\,d is completely missing. This may well be an aspect of
their data pre-processing and the difference between it and the current msMAP processing
\citep{2014PASP..126..100S}.

Other methods have been used to find flares in \kepler and other data sets.
\citet{2012ApJ...754....4O} use a statistic based on the ratio of the relative flux to noise
between adjacent bins to search {\it Hubble Space Telescope} data for flares.
\citet{2013ApJS..209....5S} detect flares in \kepler data by calculating the distribution of
brightness variations between consecutive points and selecting only those points with values three
times the value of the top 1\% of the distribution. \citet{Kowalski:2009} searched for flares
in {\it SDSS} data making use of flux changes in two photometric bands. None of these methods makes
use of the flare shape as ours does, and are therefore both likely to be less efficient. The
second method was specifically designed to look for very large flares, so it was not designed
to be efficient at finding small events.

\subsubsection{Computational time}

Another useful comparison is the computational time of the algorithms. We have coded up both
methods in {\tt python}\footnote{The code is freely available as release {\tt v1.0.0} of the {\tt
bayesflare} package \url{http://github.com/BayesFlare/bayesflare/releases}.}, with the core of our
method (Appendix~\ref{app:margamp}) written in {\tt C}. Running both algorithms for a single \kepler
Quarter 1 light curve on an Intel Core Duo 3\,GHz machine shows an average time of 6.5\,s for our
algorithm compared to the far quicker time of 165\,$\mu$s for the simpler algorithm. Despite the far
greater computational cost of our algorithm it is still fast enough that a large number of light
curves can be analysed in a reasonable time ($> 10\,000$ per day on a single machine). Our algorithm
can also make use of parallelization on multi-core machines to speed up the odds ratio calculation.

\subsection{Different flare morphologies}\label{sec:shapes}

Our flare model assumes that flares are well characterized by a fast Gaussian rise and
exponential decay shape, with time-scales of $\gtrsim 30$\,min. However, this does not describe all
flare morphologies or situations in which there are very short flares, or overlapping flares. We
have tested our algorithm on a few examples to see whether flares are still detected. We note that
this is not an exhaustive test of all potentially flare type, but does point to areas in which our
model fails or could be improved.

In the future a more complex model, or set of models, for flares could be developed. If
multiple flare models are used then the Bayesian evidence values that are calculated in our method
provide a natural way in which to discriminate between these different models. If the algorithm
currently identifies flares that do not conform to the expected shape as flares this also has an
effect on the parameter estimation (see Section~\ref{sec:pe}), as the recovered parameters will be
biased.

\subsubsection{Short flares}

The noise models we have used (see Section~\ref{sec:noisetransients}), which count short
impulsive events as noise, and the time-scale parameter ranges we use, mean that we are likely to
miss flares that evolve on short time-scales. To check the shortest flares that we are likely to
miss we have simulated sets of flares with a range of exponential decay time-scales from 5
min to 1 h at the sample rate of \kepler short cadence data (i.e.\ 30 times more samples than
long cadence data). Each of these simulations has been down-sampled to the long cadence rate by
averaging consecutive stretches of 30 data points, and then run through our algorithm. Using a log
odds ratio detection threshold of 16.5 we find an efficiency of $\roughly 12\%$ for
flares with 5 min decay time-scales, which rises to 100\% for flares with 25 min decay time
scales. This is not unexpected, but it is important to acknowledge when presenting results that
short flares will be missed.

\subsubsection{Superimposed flares}

\begin{figure}
 \begin{tabular}{c}
 \subfigure[][]{
 \label{subfig:super}
 \includegraphics[width=0.5\textwidth]{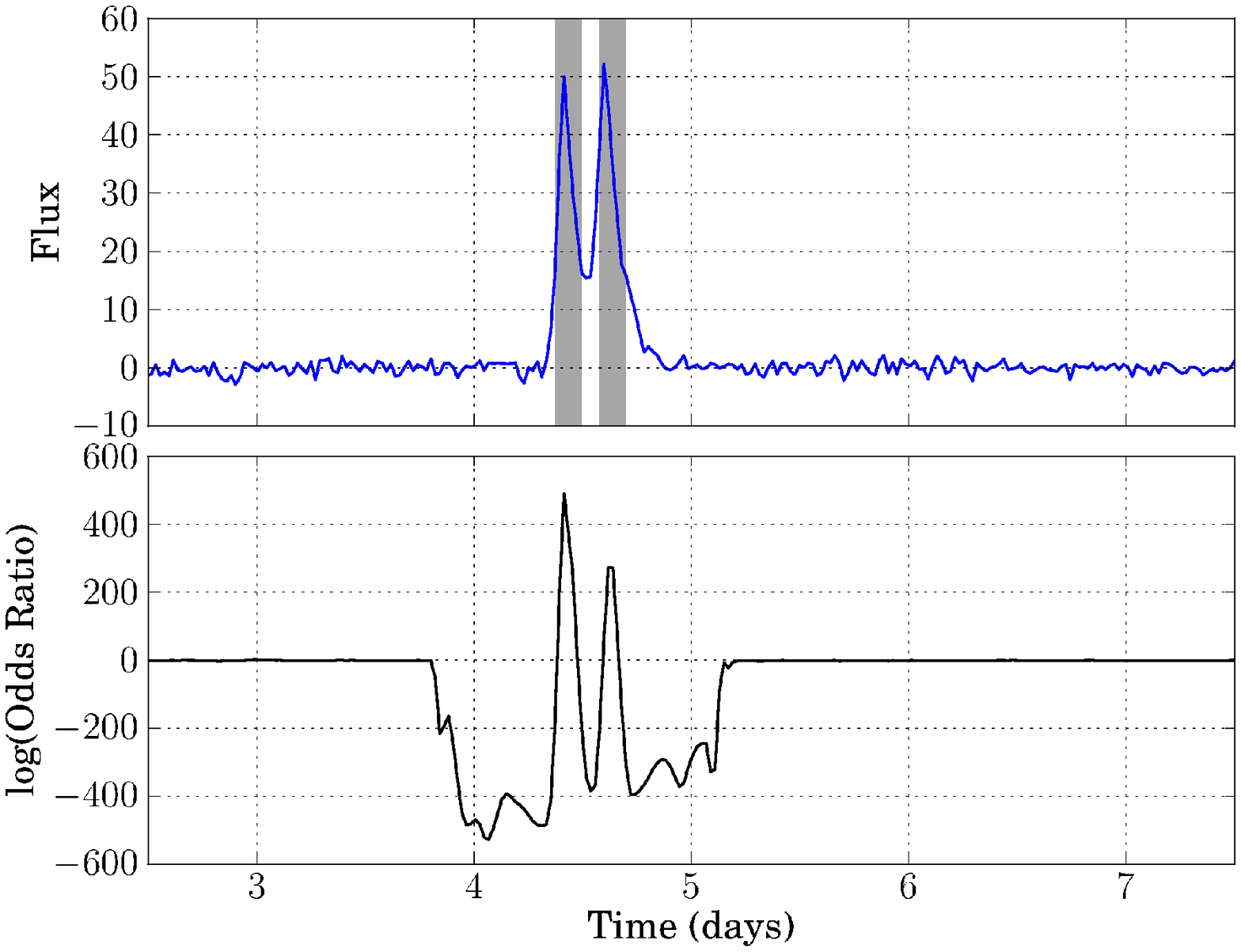}
 } \\
 \subfigure[][]{
 \label{subfig:gaussian}
 \includegraphics[width=0.5\textwidth]{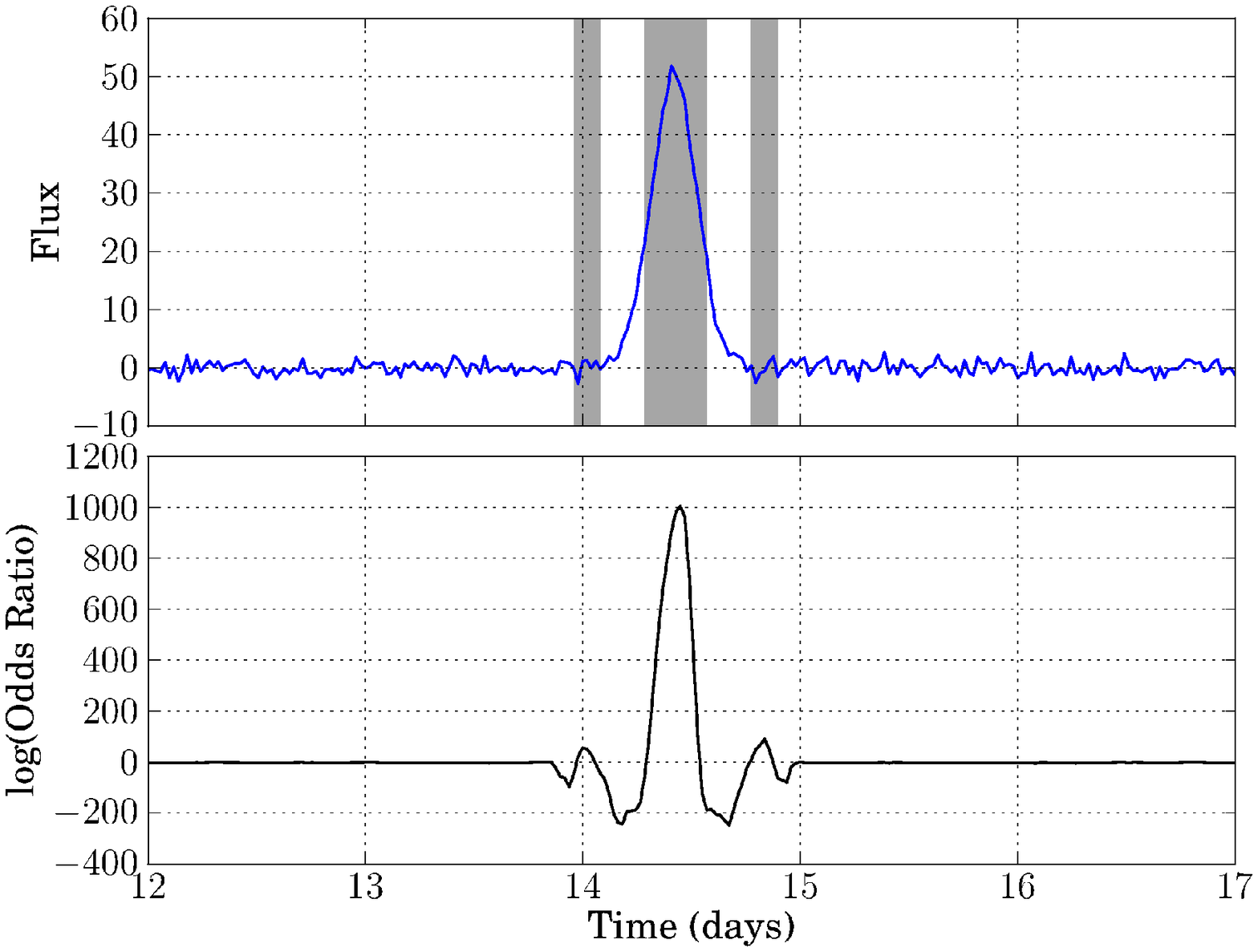}
 } \\
\subfigure[][]{
 \label{subfig:flattop}
 \includegraphics[width=0.5\textwidth]{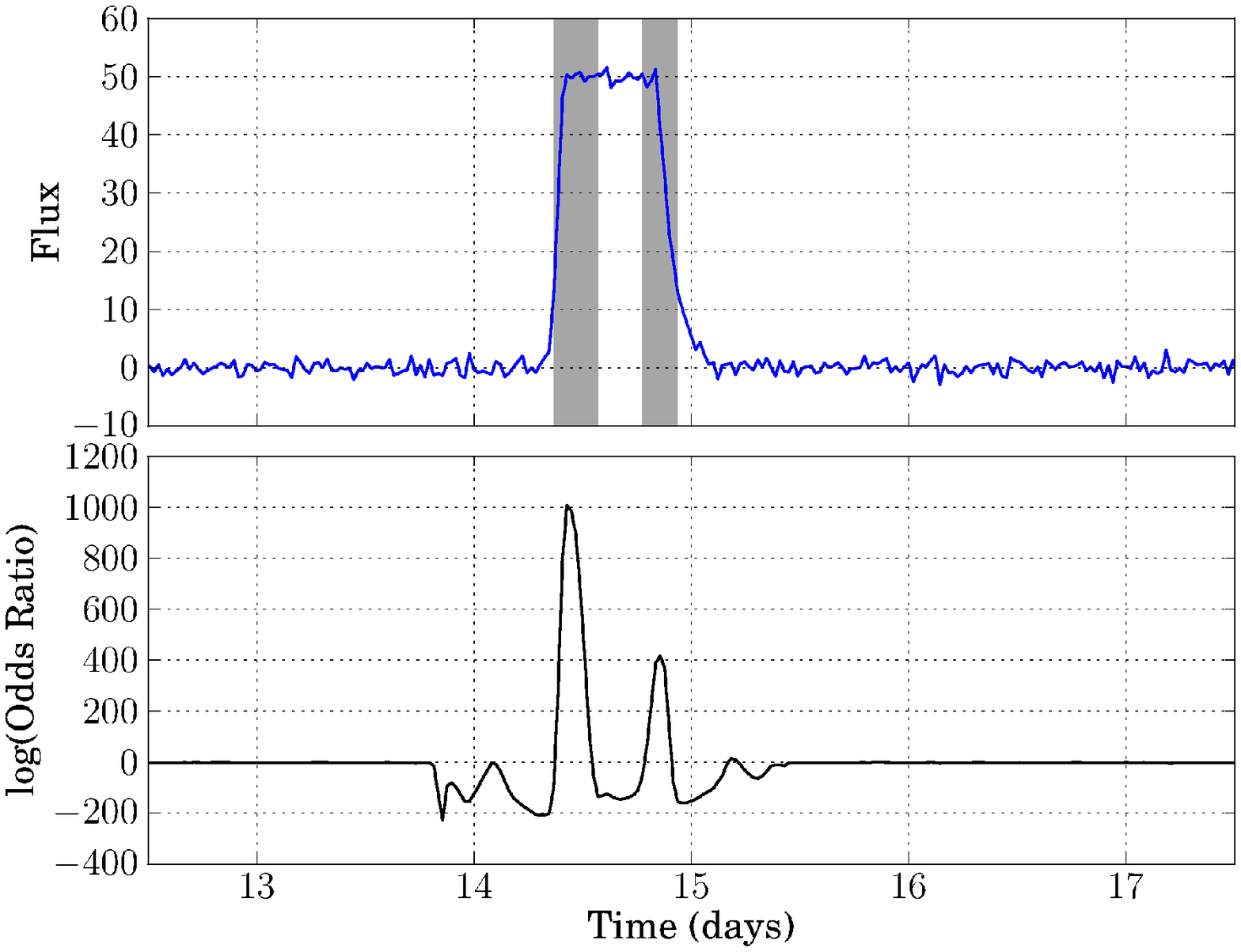}
 }
\end{tabular}
\caption{\label{fig:shapes} Examples of the algorithm run with different flare morphologies:
\subref{subfig:super} a superposition of two nearby flares, \subref{subfig:gaussian} a Gaussian
flare profile, and \subref{subfig:flattop} a flare with a flat top.}
\end{figure}

Our algorithm assumes that each analysis segment contains one flare. However, flares may
be close together and superimposed over each other. Therefore, it is interesting to see whether
superimposed flares would be detected and at what point can we tell the flares apart. We have
generated two large flares with equal time-scales and amplitudes, but with a set of peak times that
shifts one of the flares with respect to the other between 1 h and 30 h apart. We find that
the algorithm detects the two flares as a single flare when they are less than 4 h apart, but
after that can distinguish the two flares as separate. An example of this is shown in
Figure~\ref{subfig:super}, where the two flares are separated by 4.5 h. If instead we make the
second flare five times smaller in amplitude than the first flare we find that only one flare is
found until they are separated by 29 h. This is due to the noise models that are meant to
deal with artefacts generated by the larger flare are swamping the evidence for the smaller flare
when it is nearby. This shows that flares of similar amplitude can be discriminated relatively
easily, whilst it gets harder as the amplitude ratio between them increases.

\subsubsection{Gaussian signals}

One of the most generic signal shapes for a transient outburst would be a Gaussian profile.
We have simulated Gaussians of different widths ($\tau_g$) ranging from 30 min to 7 h. For
$\tau_g \lesssim 4.5$\,h the algorithm detects Gaussians as flares, but does not detect longer
Gaussians. The longer Gaussian will not be seen as they start to be fitted out by our polynomial
background model. However, for Gaussians with $\tau_g$ between approximately 2 and 3 h the
algorithm identifies the Gaussian as two or three separate flares as can be seen in
Figure~\ref{subfig:gaussian}. This is due to the signal creating extra artefacts in the detection
algorithm that are not being vetoed by our noise models.

\subsubsection{Flat-topped flares}

A final flare shape that we consider is based on the Gaussian rise and exponential decay
model, but with a flat peak that varies in width. We have generated such flares with fixed rise and
decay times, but with the flat peak lasting between 15 min and 12 h. All these signals are
detected, but often identified as two or three separate flares, as in Figure~\ref{subfig:flattop}.
It can be seen that the rise and fall appear to be identified as separate flares.

\section{Parameter estimation}\label{sec:pe}

Our analysis method also naturally lends itself to being used for estimation of the flare
parameters. However, we note that this will generally only provide reliable parameter
estimates if the flare can be well characterized by our model. In calculating the odds ratio we
marginalized over the flare amplitude, and rise and decay time-scales. However, for parameter
estimation we want to get the probability distributions of these parameters. For the amplitude,
which was analytically marginalized over, this requires a slight change in the algorithm as
described in Appendix~\ref{app:margpe}, as we still want to marginalize over the unknown background
variation. With this change we can then just evaluate the signal posterior probability, $p(A_0,
\tau_g, \tau_e, T_0|d)$, (removing the integral from equation~\ref{eq:unmarged} gives an equation
which is directly proportional to the posterior probability), over a grid in $A_0$, $\tau_g$,
$\tau_e$ and $T_0$ around the peak time corresponding to any recovered flare candidate. For
parameter estimation we relax the prior stating that $\tau_e > \tau_g$ and just set a flat prior
over the whole $\tau_e-\tau_g$ area (the black rectangle in Fig.~\ref{fig:mismatch}). Once the
posterior probability for the whole parameter volume has been calculated the posterior probability
distributions for individual parameters are obtained by marginalising over the other parameters,
e.g.\
\begin{equation}
 p(A_0|d) = \int \int \int p(A_0, \tau_g, \tau_e, T_0|d) {\rm d}\tau_g {\rm d}\tau_g {\rm d}T_0.
\end{equation}
These integrals are performed numerically with the trapezium rule. From these distributions the
most probable parameter values and credible intervals can be found, or from the posterior volume
the joint maximum a posteriori probability (i.e.\ the global maximum) for all parameters can be
found.

When analysing our flare candidates we have used a grid spanning $0 \leq \tau_g \leq 2$\,h, $0
\leq \tau_e \leq 5$\,h, an amplitude from 0 to twice the maximum dynamic range in the data (for
the 55 bins surrounding the flare), and a flare peak time ($T_0$) window of an hour either side of
the recovered peak log odds ratio time.

\begin{figure*}
 \includegraphics[width=\textwidth]{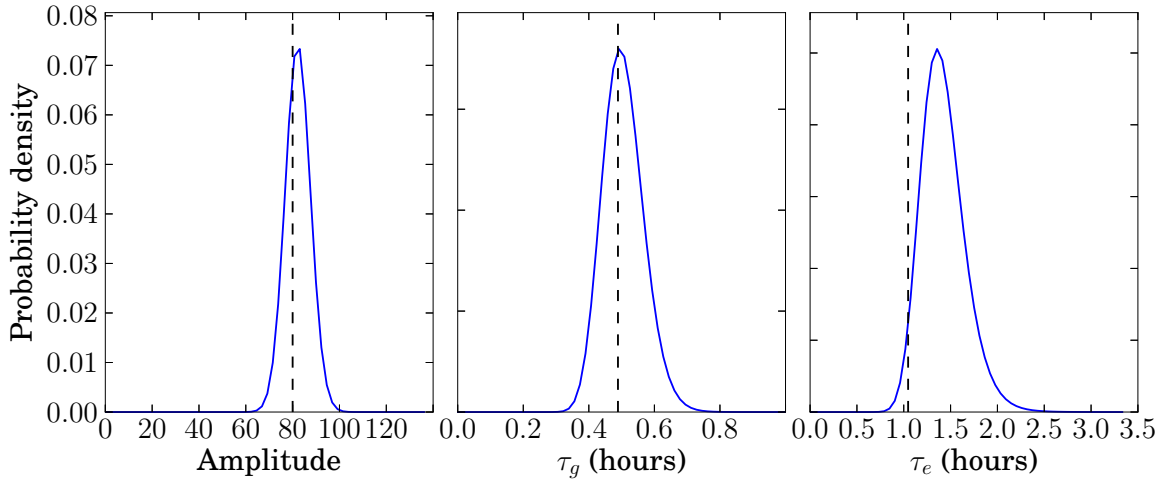}
 \caption{\label{fig:pe_sim} The parameter posterior probability distributions for a simulated
signal added to data from star KID 893676. The true values of the simulated signal are given as
vertical dashed lines.}
\end{figure*}

An example of this for a simulated flare can be seen in Fig.~\ref{fig:pe_sim}, where the true
flare parameters were $A_0 = 80$\,\fluxunits, $\tau_g = 0.49$\,hours and $\tau_e = 1.05$\,hours.
This simulation was added to real \kepler data for the star with KID 893676. The noise for this
light curve was estimated to be 6.0\,\fluxunits, giving a simulated \snr of 19.1. The
maximum a posteriori parameters that were recovered were $A_0 = 83$\,\fluxunits, $\tau_g =
0.49$\,h and $\tau_e = 1.36$\,h, giving a recovered \snr of 21.3 consistent with the
expected value. In this case as the flare time is precisely known the $T_0$ value was held fixed at
this known value. The recovered probability distribution on $\tau_e$, whilst still consistent with
the known value, does appear skewed towards high values. This is partially due to larger
values of $\tau_e$ being strongly correlated with the polynomial background coefficients. These
correlations spread the marginalized $\tau_e$ distribution towards higher values (see also the
discussion in Section~\ref{sec:res2}). These correlations could be greatly reduced by using longer
data windows (greater than the 55 time bins used here), so that the flare and the background are
more easily separated.

\begin{figure*}
 \includegraphics[width=\textwidth]{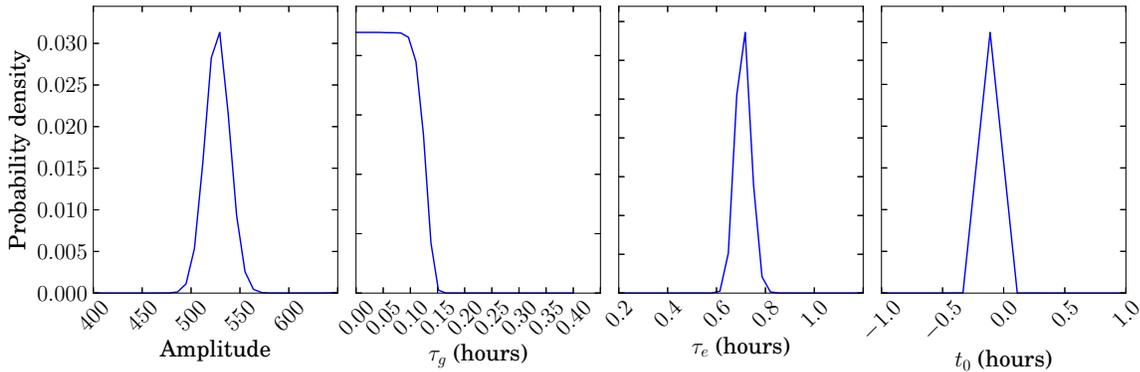}
 \caption{\label{fig:pe_real} The parameter posterior probability distributions for a real detected
signal in data from star KID 8376893.}
\end{figure*}

\begin{figure}
 \includegraphics[width=0.45\textwidth]{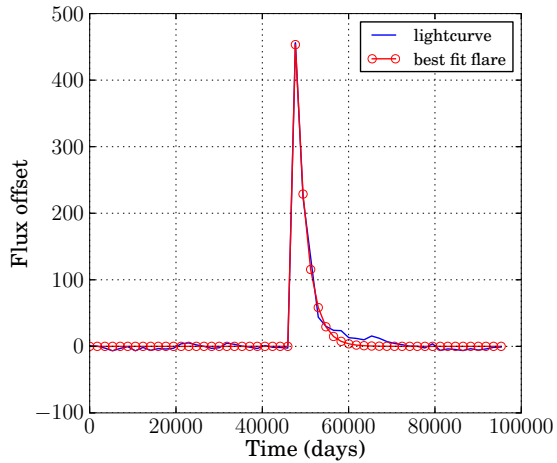}
 \caption{\label{fig:pe_bestfit} The best fit recovered flare model overlaid on the real data for
star KID 8376893.}
\end{figure}

Another example of parameter estimation, in this case for a real flare, is shown in
Fig.~\ref{fig:pe_real}. This flare was found on star KID 8376893. In this case the peak time of
the flare has also been estimated using a grid in times over a 2 h interval around the time of
the maximum odds ratio. The maximum a posteriori recovered values are $A_0 = 529$\,\fluxunits,
$\tau_g = 0$\,h and $\tau_e = 0.72$\,h, which with the estimated noise of 9.3\,\fluxunits
gives a recovered \snr of 56.2. The flat posterior probability distribution for $\tau_g$ is due to
the fact that given the size of the data time steps small values of $\tau_g$ produce
indistinguishable models. The best fit flare model is overlaid on the real data (after the removal
of the median offset) in Fig.~\ref{fig:pe_bestfit}.

\section{Analysis results}\label{sec:res}

As an initial test of the algorithm we have applied it to \kepler Quarter 1 (Q1) long-cadence data
as gathered from the public data release on the {\it Mikulski Archive for Space Telescopes}
(MAST)\footnote{\url{http://archive.stsci.edu/kepler/}}. We use the same selection criteria as used
in \citet{2011AJ....141...50W} to pick stars ($\log{g} \geq 4.2$ and $T_{\rm eff} \leq 5150$), which
returned 23\,301 stars. We veto stars that: have known transits, or are \kepler objects of
interest; are in eclipsing binaries\footnote{See e.g.\ the table at
\url{http://keplerebs.villanova.edu/}.} \citep{2011AJ....141...83P, 2012AJ....143..123M} [note that
\citet{2011AJ....141...83P} contains 33 eclipsing binaries that are not flagged in the condition
flags returned by the MAST search]; have periods of less than two days (where we take periods from
tables 1 and 2 of \citealp{2014ApJS..211...24M} and periods and secondary periods from
\citealp{2013A&A...560A...4R}); are possible red giants; or, exhibit any potential
artefacts\footnote{In practice we veto any star for which the MAST/\kepler condition flag is
defined (e.g.\ it is not `None') \url{http://archive.stsci.edu/kepler/condition_flag.html}.}. This
reduced the number of stars to 21\,746.

When performing the analysis all light curves were check for data gaps of more than a single time
bin, but none were found (otherwise we would have vetoed them). For single time bin gaps in any of
the light curves (i.e.\ {\tt NaN}s in the light curve {\tt .fits} files) we linearly interpolated
across the gaps. This is required as the algorithm makes use of discrete Fourier transforms, which
required contiguous evenly spaced data.

For all the flares detected by our algorithm we have run the parameter estimation routine (see
Section~\ref{sec:pe}) to estimate the most probable set of parameters. We have combined the
estimates of $\tau_g$ and $\tau_e$ to give a characteristic flare duration, which we define as the
timespan that encloses 95\% of the best fit flare power, centred around the point for which half
the flare's power is on either side.  We can also use these estimates to derive the recovered \snr
of the signals (assuming that they are well described by our flare model).

\subsection{Results}\label{sec:res2}

On the above data set the algorithm returned 898 flaring stars with a total of 2856 flares.
However, despite applying the above selection criterion we have viewed all light curves returning
flare candidates by eye to check for other anomalies. In doing this we have vetoed a further 160
stars, many showing strong periodic variations (some stars had high-frequency variations that led
to tens of flare signals being produced), or non-white high-frequency noise, or other oddities
(e.g.\ data discontinuities/offsets, nova-like light curves, eclipses). We note that for some stars
with detected flares the dynamic range of the underlying light curve variations was so large that
visual identification of flares was very difficult (the algorithm may do better than the eye in
this case), so in general these stars were vetoed. Other areas where the algorithm more easily
identified flares than visual inspection were cases when flare lay on steep parts of any underlying
variations.

From the parameter estimation we found that for some of the flare candidates the parameter estimates
peaked at the upper edge of the flare time-scale ranges, i.e.\ we are not including the best fitting
value within the range of our parameter estimation grid. For many flares it could be due to
correlations between our polynomial background variation model and the flare time
scales\footnote{With our current analysis window length around a flare of 27 h, flares
with $\tau_e \gtrsim 5$\,h do not decay to zero by the end of the window. This causes large
correlations with the polynomial background that heavily bias the $\tau_e$ estimate towards large
values. Longer window lengths could be used, but at the expense of a fourth order polynomial
not properly fitting the background variations for stars with higher frequency periodicities.
Further studies are needed to investigate this more thoroughly.}, but for others it may be that the
candidates just do not match the flare model very well. However, with these estimates we can
further veto flare candidates where this happens [i.e.\ veto flares for which $p(\tau_e)_{\rm max}
= 5$\,h and/or $p(\tau_g)_{\rm max} = 2$\,h]. This will bias us against long duration flares.

From this analysis, and after the above vetoing, we found 687 stars with a total of 1873 flares
\citep[cf.\ 373 star with 2358 flares in][]{2011AJ....141...50W}. Prior to the time-scale parameter
estimate veto the numbers we found were 738 stars and 2357 flares. We find flares on 305 of the 373
stars identified by \citet{2011AJ....141...50W} (of the 68 not found in our search 41 of those were
missed through our veto criteria) and on these star find on average $\roughly 70\%$ of the number of
flares they found. This means that we have confidently recovered flares on 382 additional stars than
previously found.  As we are mainly interested in the performance of our algorithm we have just
quoted numbers of flares returned by the algorithm, and we therefore do not include flares that
could be identified by subsequent visual inspection of the light curves. Inspection would reveal
more flares on the identified flaring stars.

We have visually inspected the light curves and log odds ratio time series' for the 27 unvetoed
stars where we found no candidates, but \citet{2011AJ....141...50W} did. We have also run these
through our implementation of the standard deviation threshold detection algorithm (see
Section~\ref{sec:comp}). Of these 27 only four return a flare candidate using that algorithm, which
suggests that the differently processed data sets between our analysis and theirs, and the different
noise estimation method, can make a significant difference to the results. For one of these stars
the flare is missed as our algorithm removes odds ratios for the beginning and end of the data, and
a flare fell into this period. For the other stars there appear to be either: candidate flares just
below the log odds ratio-threshold; times where potential flare-like signals are vetoed by a strong
consistency with the background model (i.e.\ the events are too short and look more like a short
transient, or are inconsistent with our flare model in some other way); or, in some cases no obvious
flares were present at all.

\begin{figure}
 \includegraphics[width=0.475\textwidth]{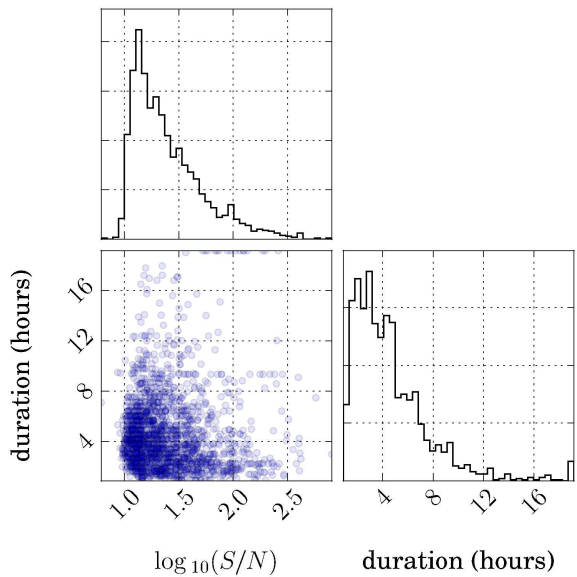}
 \caption{\label{fig:snrvsduration} The distributions of the recovered \snr and durations of
flares found in the analysis. The figure was produced using an edited version of the python {\tt
triangle} module from \url{https://github.com/dfm/triangle.py}.}
\end{figure}

Median parameters for the flare candidates after the parameter veto are given in
Table~\ref{tab:results}. The distributions for all flares in both \snr and flare duration are given
in Fig.~\ref{fig:snrvsduration}. The mean \snr is 36 and the standard deviation of the
distribution is 50, with a median value of 20, peak at $\roughly 14$, and range between 6 and 808.
The \snr distribution shows the expected fall-off to low values given our efficiency curve in
Fig.~\ref{fig:efficiency}, whilst the fall-off after the peak is related to the true distribution
of flares. The duration distribution has a mean of 4.6 h and standard deviation of 3.1 h, a
median duration of 3.8 h, a peak at 2.9 h and a range between 0.9 and 19.2 h. It is interesting
that there is a anti-correlation between the \snr and the duration. Further simulations are needed
to see if this is an effect of the algorithm or a feature of the population. However, it is hard to
see why longer high \snr signals would be missed, so it is more likely related to the population of
flares. In the future it would be worthwhile to also estimate the relative flare energies using,
for example, the method given in \citet{2011AJ....141...50W} or \citet{2013ApJS..209....5S}.

\begin{table*}
 \begin{minipage}{126mm}
  \caption{Statistics of detected flares.}
  \label{tab:results}
\begin{tabular}{lccccccc}
\hline
\kepler ID & No.\ of & $T_{\rm eff}$ & $\log{(g)}$ & Median duration & Median & Median amp. &
$\sigma$
\\
~ & flares & (K) & (cm\,s$^{-2}$) & (h) & \snr & ($e^-/{\rm s}$) & ($e^-/{\rm s}$) \\
\hline
1569863 & 4 & 3809 & 4.45 & 3.14 & 11.01 & 55.79 & 5.9 \\
1570924 & 3 & 4923 & 4.55 & 2.99 & 64.85 & 3774.86 & 59.5 \\
1722506 & 2 & 4270 & 4.43 & 12.20 & 68.62 & 963.49 & 9.8 \\
1873543 & 2 & 3702 & 4.52 & 4.33 & 25.76 & 176.47 & 8.0 \\
2013754 & 5 & 4280 & 4.57 & 4.55 & 12.06 & 79.42 & 7.8 \\
2140782 & 1 & 4946 & 4.62 & 2.35 & 12.87 & 77.39 & 5.1 \\
2437317 & 1 & 4250 & 4.70 & 7.49 & 575.65 & 2574.92 & 8.6 \\
2441562 & 1 & 3806 & 4.46 & 4.49 & 37.02 & 149.61 & 5.0 \\
2442866 & 3 & 4344 & 4.49 & 9.44 & 14.56 & 62.76 & 8.6 \\
2570846 & 1 & 3865 & 4.61 & 2.62 & 24.98 & 152.08 & 5.3 \\
\hline
\end{tabular}
\medskip
This is an abridged version of the table. A full version is available with the online version of
the article.
 \end{minipage}
\end{table*}

We do find a few cases where an obvious loud flare is missed. An example is KID 8176468 where five
flares are found, but not the loudest looking event (at around 25.5 d into the data set).
Looking at this case we see that this flare is strongly consistent with the background model,
because even though it is loud it actually has a decay time that is shorter than our model range.
This is consistent with our knowledge that we will be biased against short flares, but in this
particular case, where the flare is obvious and has a long tail, it may suggest that the flare model
we use may need augmenting to allow a steeper initial decay.

\begin{figure*}
 \includegraphics[width=\textwidth]{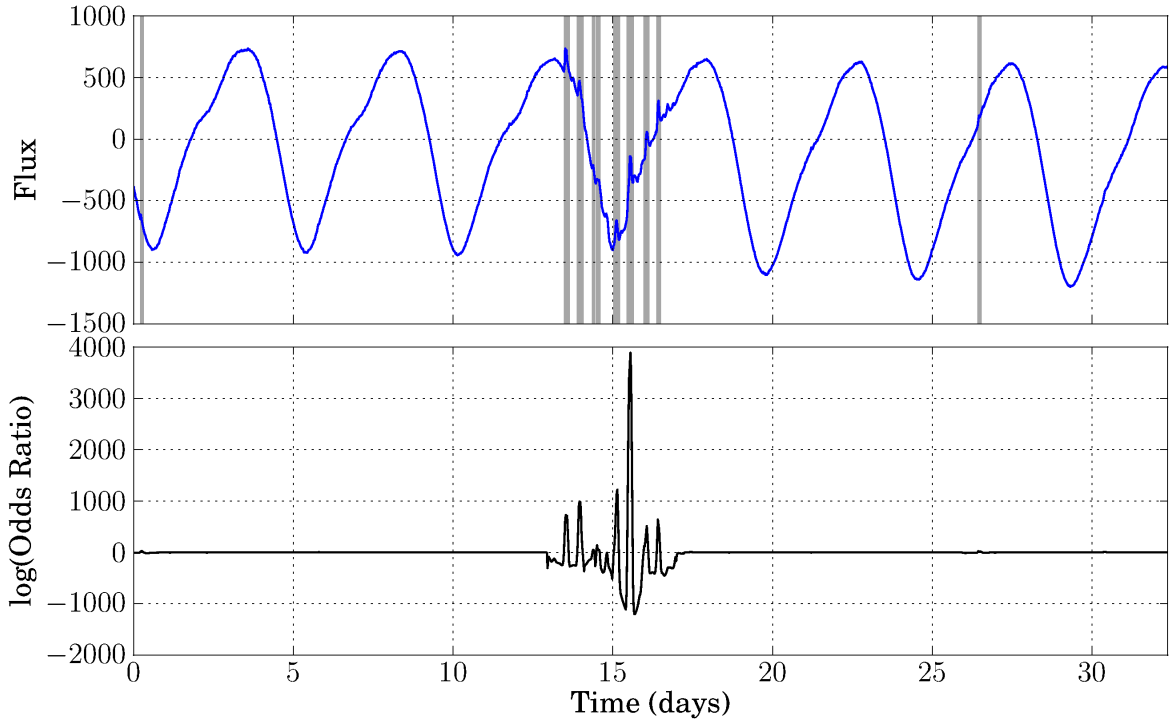}
 \caption{\label{fig:interesting} The light curve and log odds ratio time series for KID 9450669
showing an intense period of flaring activity. Shaded areas in the upper plot represent times when
the log odds ratio exceeds our threshold of 16.5.}
\end{figure*}

Visual inspection of the flaring light curves identified some interesting candidates. A
particularly noteworthy example was KID 9450669, which appeared to undergo an intense period of
flaring activity before dropping back to a more normal rate. The light curve and odds ratio time
series can be seen in Fig.~\ref{fig:interesting}. Inspection of light curves from all later
\kepler Quarters does not show such an intense period of activity again.

\section{Conclusions}

This paper has been primarily focused on the development and characterization of an algorithm for
detecting flares. The algorithm works under the general assumption that flares have a
characteristic shape that we model as a Gaussian rise and exponential decay. This assumption seems
reasonable from observations of flares, but a more generic model (or indeed a more specific
model if there is good theoretical motivation for a particular flare shape) could in the
future prove more appropriate. Along with the flare model we have assumed that the data contains
unknown low frequency variations, which we model as a fourth order polynomial. The algorithm
computes an odds ratio comparing the probability of the data containing a flare model plus the
background polynomial variations (the signal model) to the probability of the data just consisting
of the background variations {\it or} the background variations plus very short duration transients
within the analysis window (the background model). This method allows further signal or background
models to be added in the future based on updated knowledge of flare morphologies, or better
understanding of background artefacts. Another fundamental assumption is that the noise in the
data can be modelled with a Gaussian likelihood, which we have seen in Section~\ref{sec:real} is
not wholly reliable, but can be robust provided a conservative threshold is set.

We have characterized this method using simulated data sets consisting of Gaussian white noise plus
a low frequency sinusoidal background. If requiring a FAP of 0.1\% (based on data that are the
length of \kepler Quarter 1 long cadence data) from noise alone this provides a detection threshold
on the log odds ratio of 8.3. When this threshold is used for simulated data sets with fake signals
of varying parameters and \snr it returns a 95\% detection efficiency for signals with \snr $\gtrsim
13$. However, we find that when adding signals they can occasionally produce spurious extra false
detections at times when the analysis window is sliding on to the flare. To
reduce these false alarms to occur for only $\roughly 1$\% of simulated flares requires increasing
the detection threshold to 16.5. This corresponds to a 95\% detection efficiency for signals with
\snr $\gtrsim 20$. Studies using real \kepler data for ``quiet'' stars, in which the noise
deviates from the Gaussian assumption, show that this threshold of 16.5 again gives an
approximately 1\% false alarm rate.

Applying our method to a selection of stars within the ranges $\log{g} \geq 4.2$ and $T_{\rm eff}
\leq 5150$ in \kepler Q1 data (see Section~\ref{sec:res} for other veto criteria) we find 687 stars
exhibiting flares \citep[cf. 373 stars in the analysis of][]{2011AJ....141...50W}. We have not
aimed at attempting a detailed statistical analysis of distributions of flare properties, but have
shown basic distributions of flare duration and \snr.

Future modifications to the algorithm could be to include extra models for the background
variations that would be a better fit to the low frequency variations than the polynomial
currently used. This could be done by adding a low-frequency sinusoid model with unknown amplitude,
frequency and phase to the current polynomial fit. The unknown amplitude and phase could be
analytically marginalized over by splitting $m = A\sin{(2\pi ft + \phi)}$ into $m = B\sin{(2\pi ft)}
+ C\cos{(2\pi ft)}$ and using the algorithm described in Appendix~\ref{app:margamp}. However, we
would need to numerically search of this frequency range, which would slow down the method.
Such a method could allow longer windows to be used and potentially reduce the additional artefacts
caused by flares themselves, meaning that non-flare-like transients (such as transits and
eclipses) would not trigger a flare detection. This could allow the threshold to be significantly
lowered allowing smaller flares to be recovered. As discussed in Section~\ref{sec:real}
further modifications to the noise model to account for short period correlations seen in real
\kepler data could also be taken into account. Another type of artefact that triggered the
algorithm and was observed during visual inspection of \kepler data was large step offsets between
stretches of data. This type of artefact could easily be modelled in the noise model by including a
step function of unknown amplitude (to be analytically marginalized over). The algorithm currently
relies on data being contiguous over a \kepler Quarter. However, future quarters contain some large
gaps. These can easily be dealt with by making the algorithm treat data separated by large gaps as
independent analyses.

We plan to run the algorithm on further \kepler Quarters to build up a complete picture of flaring
activity at statistics on these stars. This would include running on short-cadence data. In
further studies it would also be interesting to compare any times highlighted by our method as
strongly favouring the noise model with \kepler data quality flags.  We also
expect our method to be applicable to other flare searches, in particular using the large sets of
multi-wavelength solar observations.

\section*{Acknowledgements}
DW has been funded for this work by the Royal Society of Edinburgh Cormack Bequest. MP is
funded by STFC under grant ST/L000946/1. LF acknowledges support from STFC grants ST/I001808 and
ST/L000741. We are very grateful to the referee for comments that have helped greatly improve
the paper. We would like to thank NASA and the \kepler mission team for the data we have used.

\bibliography{flare_detection}

\begin{thebibliography}{}
 \providecommand{\href}[2]{#2}
  \providecommand{\doi}[1]{\href{http://dx.doi.org/#1}{doi:#1}}
  \providecommand{\eprint}[1]{\href{http://arxiv.org/abs/#1}{arXiv:#1}}

\bibitem[\protect\citeauthoryear{{Balona}}{{Balona}}{2012}]{2012MNRAS.423.3420B}
{Balona} L.~A.,  2012, \mnras, 423, 3420

\bibitem[\protect\citeauthoryear{{Basri} et~al.,}{{Basri}
  et~al.}{2011}]{2011AJ....141...20B}
{Basri} G.  et~al., 2011, \aj, 141, 20, \eprint{1008.1092}

\bibitem[\protect\citeauthoryear{{Basri} et~al.,}{{Basri}
  et~al.}{2010}]{2010ApJ...713L.155B}
{Basri} G.  et~al., 2010, \apjl, 713, L155, \eprint{1001.0414}

\bibitem[\protect\citeauthoryear{{Batalha} et~al.,}{{Batalha}
  et~al.}{2013}]{2013ApJS..204...24B}
{Batalha} N.~M.  et~al., 2013, \apjs, 204, 24, \eprint{1202.5852}

\bibitem[\protect\citeauthoryear{{Benz} \& {G{\"u}del}}{{Benz} \&
  {G{\"u}del}}{2010}]{2010ARA&A..48..241B}
{Benz} A.~O.,  {G{\"u}del} M.,  2010, \araa, 48, 241

\bibitem[\protect\citeauthoryear{{Bretthorst}}{{Bretthorst}}{1988}]{Bretthorst:1988}
{Bretthorst} G.~L.,  1988, Bayesian Spectrum Analysis and Parameter Estimation.
No.~48 in Lecture Notes in Statistics, Springer-Verlag

\bibitem[\protect\citeauthoryear{{Cameron}}{{Cameron}}{2011}]{2011PASA...28..128C}
{Cameron} E.,  2011, PASA, 28, 128, \eprint{1012.0566}

\bibitem[\protect\citeauthoryear{{Carrington}}{{Carrington}}{1859}]{1859MNRAS..20...13C}
{Carrington} R.~C.,  1859, \mnras, 20, 13

\bibitem[\protect\citeauthoryear{{Christiansen} et~al.,}{{Christiansen}
  et~al.}{2013}]{KeplerDH}
{Christiansen} J.~L.  et~al., 2013, Technical Report KSCI-19040-00, Kepler Data
  Characteristics Handbook

\bibitem[\protect\citeauthoryear{{Clark}, {Heng}, {Pitkin} \& {Woan}}{{Clark}
  et~al.}{2007}]{2007PhRvD..76d3003C}
{Clark} J.,  {Heng} I.~S.,  {Pitkin} M.,    {Woan} G.,  2007, \prd, 76, 043003,
  \eprint{gr-qc/0703138}

\bibitem[\protect\citeauthoryear{{Dal} \& {Evren}}{{Dal} \&
  {Evren}}{2012}]{Dal:2012}
{Dal} H.~A.,  {Evren} S.,  2012, New Astronomy, 17, 399, \eprint{1206.3761}

\bibitem[\protect\citeauthoryear{{Emslie} et~al.,}{{Emslie}
  et~al.}{2012}]{2012ApJ...759...71E}
{Emslie} A.~G.  et~al., 2012, \apj, 759, 71, \eprint{1209.2654}

\bibitem[\protect\citeauthoryear{{Fletcher} et~al.,}{{Fletcher}
  et~al.}{2011}]{2011SSRv..159...19F}
{Fletcher} L.  et~al., 2011, \ssr, 159, 19, \eprint{1109.5932}

\bibitem[\protect\citeauthoryear{{Gershberg}}{{Gershberg}}{1972}]{Gershberg:1972}
{Gershberg} R.~E.,  1972, \apss, 19, 75

\bibitem[\protect\citeauthoryear{{Hambaryan}, {Neuh{\"a}user} \&
  {Stelzer}}{{Hambaryan} et~al.}{1999}]{1999A&A...345..121H}
{Hambaryan} V.,  {Neuh{\"a}user} R.,    {Stelzer} B.,  1999, \aap, 345, 121

\bibitem[\protect\citeauthoryear{{Hannah}, {Hudson}, {Battaglia}, {Christe},
  {Ka{\v s}parov{\'a}}, {Krucker}, {Kundu} \& {Veronig}}{{Hannah}
  et~al.}{2011}]{2011SSRv..159..263H}
{Hannah} I.~G.,  {Hudson} H.~S.,  {Battaglia} M.,  {Christe} S.,  {Ka{\v
  s}parov{\'a}} J.,  {Krucker} S.,  {Kundu} M.~R.,    {Veronig} A.,  2011,
  \ssr, 159, 263, \eprint{1108.6203}

\bibitem[\protect\citeauthoryear{{Hilton}, {West}, {Hawley} \&
  {Kowalski}}{{Hilton} et~al.}{2010}]{Hilton:2010}
{Hilton} E.~J.,  {West} A.~A.,  {Hawley} S.~L.,    {Kowalski} A.~F.,  2010,
  \aj, 140, 1402, \eprint{1009.1158}

\bibitem[\protect\citeauthoryear{{Hudson}}{{Hudson}}{1991}]{1991SoPh..133..357H}
{Hudson} H.~S.,  1991, \solphys, 133, 357

\bibitem[\protect\citeauthoryear{{Hudson}, {Wolfson} \& {Metcalf}}{{Hudson}
  et~al.}{2006}]{2006SoPh..234...79H}
{Hudson} H.~S.,  {Wolfson} C.~J.,    {Metcalf} T.~R.,  2006, \solphys, 234, 79

\bibitem[\protect\citeauthoryear{{Ishida}, {Ichimura}, {Shimizu} \&
  {Mahasenaputra}}{{Ishida} et~al.}{1991}]{Ishida:1991}
{Ishida} K.,  {Ichimura} K.,  {Shimizu} Y.,    {Mahasenaputra} 1991, \apss,
  182, 227

\bibitem[\protect\citeauthoryear{{Jeffreys}}{{Jeffreys}}{1998}]{Jeffreys:1931}
{Jeffreys} H.,  1998, Theory of Probability, 3 edn.
Oxford University Press

\bibitem[\protect\citeauthoryear{{Jenkins} et~al.,}{{Jenkins}
  et~al.}{2010a}]{2010ApJ...713L..87J}
{Jenkins} J.~M.  et~al., 2010a, \apjl, 713, L87, \eprint{1001.0258}

\bibitem[\protect\citeauthoryear{{Jenkins} et~al.,}{{Jenkins}
  et~al.}{2010b}]{2010ApJ...713L.120J}
{Jenkins} J.~M.  et~al., 2010b, \apjl, 713, L120, \eprint{1001.0256}

\bibitem[\protect\citeauthoryear{{Kowalski}, {Hawley}, {Hilton}, {Becker},
  {West}, {Bochanski} \& {Sesar}}{{Kowalski} et~al.}{2009}]{Kowalski:2009}
{Kowalski} A.~F.,  {Hawley} S.~L.,  {Hilton} E.~J.,  {Becker} A.~C.,  {West}
  A.~A.,  {Bochanski} J.~J.,    {Sesar} B.,  2009, \aj, 138, 633,
  \eprint{0906.2030}

\bibitem[\protect\citeauthoryear{{Kowalski}, {Hawley}, {Holtzman}, {Wisniewski}
  \& {Hilton}}{{Kowalski} et~al.}{2010}]{2010ApJ...714L..98K}
{Kowalski} A.~F.,  {Hawley} S.~L.,  {Holtzman} J.~A.,  {Wisniewski} J.~P.,
  {Hilton} E.~J.,  2010, \apjl, 714, L98, \eprint{1003.3057}

\bibitem[\protect\citeauthoryear{{Kowalski}, {Hawley}, {Wisniewski}, {Osten},
  {Hilton}, {Holtzman}, {Schmidt} \& {Davenport}}{{Kowalski}
  et~al.}{2013}]{2013ApJS..207...15K}
{Kowalski} A.~F.,  {Hawley} S.~L.,  {Wisniewski} J.~P.,  {Osten} R.~A.,
  {Hilton} E.~J.,  {Holtzman} J.~A.,  {Schmidt} S.~J.,    {Davenport} J.~R.~A.,
   2013, \apjs, 207, 15, \eprint{1307.2099}

\bibitem[\protect\citeauthoryear{{Kowalski}, {Mathioudakis}, {Hawley},
  {Hilton}, {Dhillon}, {Marsh} \& {Copperwheat}}{{Kowalski}
  et~al.}{2011}]{2011ASPC..448.1157K}
{Kowalski} A.~F.,  {Mathioudakis} M.,  {Hawley} S.~L.,  {Hilton} E.~J.,
  {Dhillon} V.~S.,  {Marsh} T.~R.,    {Copperwheat} C.~M.,  2011, in
  {Johns-Krull} C.,  {Browning} M.~K.,   {West} A.~A.,  eds,  Astronomical
  Society of the Pacific Conference Series Vol. 448, 16th Cambridge Workshop on
  Cool Stars, Stellar Systems, and the Sun. p.~1157, \eprint{1103.0822}

\bibitem[\protect\citeauthoryear{{Lacy}, {Moffett} \& {Evans}}{{Lacy}
  et~al.}{1976}]{Lacy:1976}
{Lacy} C.~H.,  {Moffett} T.~J.,    {Evans} D.~S.,  1976, \apjs, 30, 85

\bibitem[\protect\citeauthoryear{{Maehara} et~al.,}{{Maehara}
  et~al.}{2012}]{2012Natur.485..478M}
{Maehara} H.  et~al., 2012, \nat, 485, 478

\bibitem[\protect\citeauthoryear{{Matijevi{\v c}}, {Pr{\v s}a}, {Orosz},
  {Welsh}, {Bloemen} \& {Barclay}}{{Matijevi{\v c}}
  et~al.}{2012}]{2012AJ....143..123M}
{Matijevi{\v c}} G.,  {Pr{\v s}a} A.,  {Orosz} J.~A.,  {Welsh} W.~F.,
  {Bloemen} S.,    {Barclay} T.,  2012, \aj, 143, 123, \eprint{1204.2113}

\bibitem[\protect\citeauthoryear{{McQuillan}, {Aigrain} \&
  {Roberts}}{{McQuillan} et~al.}{2012}]{2012A&A...539A.137M}
{McQuillan} A.,  {Aigrain} S.,    {Roberts} S.,  2012, \aap, 539, A137,
  \eprint{1111.5580}

\bibitem[\protect\citeauthoryear{{McQuillan}, {Mazeh} \& {Aigrain}}{{McQuillan}
  et~al.}{2014}]{2014ApJS..211...24M}
{McQuillan} A.,  {Mazeh} T.,    {Aigrain} S.,  2014, \apjs, 211, 24,
  \eprint{1402.5694}

\bibitem[\protect\citeauthoryear{{Moffett}}{{Moffett}}{1974}]{Moffett:1974}
{Moffett} T.~J.,  1974, \apjs, 29, 1

\bibitem[\protect\citeauthoryear{{Moffett} \& {Bopp}}{{Moffett} \&
  {Bopp}}{1976}]{1976ApJS...31...61M}
{Moffett} T.~J.,  {Bopp} B.~W.,  1976, \apjs, 31, 61

\bibitem[\protect\citeauthoryear{{Osten}, {Kowalski}, {Sahu} \&
  {Hawley}}{{Osten} et~al.}{2012}]{2012ApJ...754....4O}
{Osten} R.~A.,  {Kowalski} A.,  {Sahu} K.,    {Hawley} S.~L.,  2012, \apj, 754,
  4, \eprint{1205.1485}

\bibitem[\protect\citeauthoryear{{Pr{\v s}a} et~al.,}{{Pr{\v s}a}
  et~al.}{2011}]{2011AJ....141...83P}
{Pr{\v s}a} A.  et~al., 2011, \aj, 141, 83, \eprint{1006.2815}

\bibitem[\protect\citeauthoryear{{Reinhold}, {Reiners} \& {Basri}}{{Reinhold}
  et~al.}{2013}]{2013A&A...560A...4R}
{Reinhold} T.,  {Reiners} A.,    {Basri} G.,  2013, \aap, 560, A4,
  \eprint{1308.1508}

\bibitem[\protect\citeauthoryear{{Roberts}, {McQuillan}, {Reece} \&
  {Aigrain}}{{Roberts} et~al.}{2013}]{2013arXiv1308.3644R}
{Roberts} S.,  {McQuillan} A.,  {Reece} S.,    {Aigrain} S.,  2013, \mnras,
  435, 3639, \eprint{1308.3644}

\bibitem[\protect\citeauthoryear{Rosner}{Rosner}{1983}]{Rosner:1983}
Rosner B.,  1983, Technometrics, 25, 165

\bibitem[\protect\citeauthoryear{{Savitzky} \& {Golay}}{{Savitzky} \&
  {Golay}}{1964}]{Savitzky:1964}
{Savitzky} A.,  {Golay} M.~J.~E.,  1964, Analytical Chemistry, 36, 1627

\bibitem[\protect\citeauthoryear{{Scargle}}{{Scargle}}{1998}]{1998ApJ...504..405S}
{Scargle} J.~D.,  1998, \apj, 504, 405, \eprint{astro-ph/9711233}

\bibitem[\protect\citeauthoryear{{Searle}, {Sutton} \& {Tinto}}{{Searle}
  et~al.}{2009}]{2009CQGra..26o5017S}
{Searle} A.~C.,  {Sutton} P.~J.,    {Tinto} M.,  2009, Classical and Quantum
  Gravity, 26, 155017, \eprint{0809.2809}

\bibitem[\protect\citeauthoryear{{Searle}, {Sutton}, {Tinto} \&
  {Woan}}{{Searle} et~al.}{2008}]{2008CQGra..25k4038S}
{Searle} A.~C.,  {Sutton} P.~J.,  {Tinto} M.,    {Woan} G.,  2008, Classical
  and Quantum Gravity, 25, 114038, \eprint{0712.0196}

\bibitem[\protect\citeauthoryear{{Shibayama} et~al.,}{{Shibayama}
  et~al.}{2013}]{2013ApJS..209....5S}
{Shibayama} T.  et~al., 2013, \apjs, 209, 5, \eprint{1308.1480}

\bibitem[\protect\citeauthoryear{{Shlens}}{{Shlens}}{2014}]{Shlens:2014}
{Shlens} J.,  2014, ArXiv e-prints, \eprint{1404.1100}

\bibitem[\protect\citeauthoryear{{Smith} et~al.,}{{Smith}
  et~al.}{2012}]{2012PASP..124.1000S}
{Smith} J.~C.  et~al., 2012, \pasp, 124, 1000, \eprint{1203.1383}

\bibitem[\protect\citeauthoryear{{Stumpe}, {Smith}, {Catanzarite}, {Van Cleve},
  {Jenkins}, {Twicken} \& {Girouard}}{{Stumpe}
  et~al.}{2014}]{2014PASP..126..100S}
{Stumpe} M.~C.,  {Smith} J.~C.,  {Catanzarite} J.~H.,  {Van Cleve} J.~E.,
  {Jenkins} J.~M.,  {Twicken} J.~D.,    {Girouard} F.~R.,  2014, \pasp, 126,
  100

\bibitem[\protect\citeauthoryear{{Stumpe} et~al.,}{{Stumpe}
  et~al.}{2012}]{2012PASP..124..985S}
{Stumpe} M.~C.  et~al., 2012, \pasp, 124, 985, \eprint{1203.1382}

\bibitem[\protect\citeauthoryear{{Thompson} et~al.,}{{Thompson}
  et~al.}{2013}]{KeplerRelease21}
{Thompson} S.~E.  et~al., 2013, Technical Report KSCI-19061-001, Kepler Data
  Release 21 Notes

\bibitem[\protect\citeauthoryear{{Walkowicz} et~al.,}{{Walkowicz}
  et~al.}{2011}]{2011AJ....141...50W}
{Walkowicz} L.~M.  et~al., 2011, \aj, 141, 50, \eprint{1008.0853}

\bibitem[\protect\citeauthoryear{{Woods} et~al.,}{{Woods}
  et~al.}{2004}]{2004GeoRL..3110802W}
{Woods} T.~N.  et~al., 2004, \grl, 31, 10802

\end{thebibliography}

\appendix

\section{Marginalizing over a polynomial background}\label{app:margamp}

In Section~\ref{sec:varback} we introduced a polynomial background variation. Rather than having to
integrate numerically over each polynomial coefficient it is possible to analytically integrate
each of them. We define a generic model as $m = Af + g$, where $f(x)$ is an arbitrary function
with an amplitude that can be factored out as $A$, and $g$ contains any other components of the
model. If we substitute this into the odds ratio defined in equation~\ref{eq:unmarged}, and place
the integral over $A$ between $[-\infty, \infty]$, we can rearrange it to give
\begin{equation}\label{eq:ampmarg}
 \mathcal{O} = \int_{-\infty}^{\infty} \exp{\left(-\frac{1}{2\sigma^2}(A^2X + AY + Z)\right)} p(A)
{\rm d}A,
\end{equation}
where $X = \sum_j f_j^2$, $Y = 2\left(\sum_j f_jg_j - \sum_j d_jf_j\right)$ and $Z = \sum_j g_j^2
-2\sum_j d_jg_j$. Provided that the prior on $A$, $p(A)$ is constant, or varies very little, over
the range in which the likelihood ratio is significant (e.g.\ is very wide Gaussian), this integral
is solved to give
\begin{equation}\label{eq:inftoinf}
 \mathcal{O} = p(A) \sqrt{\frac{2\pi\sigma^2}{X}}\exp{\left(-\frac{(Z-Y^2/4X)}{2\sigma^2}
\right)}.
\end{equation}
For the case where the amplitude $A$ must be purely positive (e.g.\ a flare) then
equation~\ref{eq:ampmarg} can be integrated between $[0, \infty]$ to give
\begin{equation}\label{eq:zerotoinf}
 \mathcal{O} = p(A) \sqrt{\frac{\pi\sigma^2}{2X}}\exp{\left(-\frac{(Z-Y^2/4X)}{2\sigma^2}
\right)\erfc{\left( \frac{Y}{2\sqrt{2\sigma^2X}} \right)}}.
\end{equation}

If we consider a model containing a third order polynomial and a flare signal
\begin{equation}
 m = A_0 m_f(t, \tau_g, \tau_e, T_0) + A + Bt + Ct^2 + Dt^3
\end{equation}
the odds ratio in equation~\ref{eq:unmarged} would instead be given by
\begin{align}
\mathcal{O}(T_0) =& \int_{\tau_g} \int_{\tau_e} \int_{0}^{\infty} \int_{-\infty}^{\infty}
\int_{-\infty}^{\infty} \int_{-\infty}^{\infty} \int_{-\infty}^{\infty} e^{-\frac{\sum
(m^2 - 2dm)}{2\sigma^2}} \times \nonumber \\
 & p(\tau_g, \tau_e, A_0, A, B, C, D){\rm d}\tau_g {\rm d}\tau_g {\rm d}A_0 {\rm d}A {\rm d}B {\rm
d}C {\rm d}D,
\end{align}
which can be analytically reduced by four applications of equation~\ref{eq:inftoinf} followed by a
final application of equation~\ref{eq:zerotoinf}. Note that the integral over $[0, \infty]$ can
only be used for one parameter and must be performed last. If explicitly performing the integral for
the five amplitudes being marginalized over above (e.g.\ using a symbolic mathematics program) the
algebraic expression is far too large to be easily transcribed into an analysis code. However, we
have developed an algorithm that can perform the integral for an arbitrary number of model component
amplitudes without the need to explicitly write out the whole expression\footnote{A version of
this algorithm is provided in the {\tt amplitude-marginalizer} function library found at
\url{http://github.com/mattpitkin/amplitude-marginalizer}}. This is given in pseudo-code below.

If performing all amplitude marginalizations between $[-\infty, \infty]$ an alternative approach to
simpifying the integral would be to create an orthogonal set of model components (via
diagonalization of a model component matrix), which would mean that all cross terms between the {\it
orthogonal} models were zero \citep[see e.g.][]{Bretthorst:1988}.

\subsection{The marginalization algorithm}\label{sec:alg}

Given $N$ model components for which an amplitude can be factored out we have a generic model
\begin{equation}
 m = \sum_i^N A_i f_i
\end{equation}
where $A_i$ are the amplitude components and $f_i$ are the component functions. We can create a
matrix, \texttt{C}, containing the sums of the products of each of the component functions with
each other and with the data $d$, such that
\begin{equation*}
 {\texttt C} =
\left(
      \begin{array}{c c c c}
      -2\sum_j d_j f_{1j} & 2\sum_j f_{1j} f_{2j} & 2\sum_j f_{1j} f_{3j} & \ldots \\
      0                   & -2\sum_j d_j f_{2j}   & 2\sum_j f_{2j} f_{3j} & \ldots \\
      0                   & 0                     & -2\sum_j d_j f_{3j} & \ldots \\
      \vdots              & \vdots                & \vdots              & \ddots
      \end{array}
\right),
\end{equation*}
where the sums are over all the model and data points. We can also create a vector containing the
squared model terms
\begin{equation*}
 {\texttt S} = \left\{\sum_j f_{1j}^2, \sum_j f_{2j}^2, \sum_j f_{3j}^2, \ldots\right\}.
\end{equation*}
Algorithm~\ref{alg:logO} can then be applied to produce the $X$, $Y$ and $Z$ values given
in equations~\ref{eq:inftoinf} and \ref{eq:zerotoinf} and calculate the log of the odds ratio
marginalized over the amplitudes of the model components.
\begin{algorithm}
\caption{Calculate $\log{\mathcal{O}}$ marginalized over model component amplitudes (this assumes
\texttt{C}-style array indexing starting from 0).}
\label{alg:logO}
\begin{algorithmic}[1]
 \STATE $Z = 0$
 \STATE $M = N-1$
 \STATE $\log{\mathcal{O}} = 0$
 \FOR{$i=0$ \TO $M-1$}
   \FOR{$j=i$ \TO $M$}
     \FOR{$k=i$ \TO $M$}
       \IF{$j$ is equal to $i$}
         \IF{$k > j$}
           \STATE $\texttt{S}[k] = \texttt{S}[k] - \texttt{C}[j][k]^2/4\texttt{S}[i]$
         \ELSIF{$k$ is equal to $j$}
           \STATE $Z = Z - \texttt{C}[j][k]^2/4\texttt{S}[i]$
         \ENDIF
       \ELSE
         \IF{$k$ is equal to $i$}
           \STATE $\texttt{C}[j][j] = \texttt{C}[j][j] - \texttt{C}[i][j]\times
\texttt{C}[i][k]/2\texttt{S}[i]$
         \ELSIF{$k > j$}
           \STATE $\texttt{C}[j][k] = \texttt{C}[j][k] - \texttt{C}[i][j]\times
\texttt{C}[i][k]/2\texttt{S}[i]$
         \ENDIF
       \ENDIF
     \ENDFOR
   \ENDFOR
 \ENDFOR
 \STATE $X = \texttt{S}[M]$
 \STATE $Y = \texttt{C}[M][M]$
 \FOR{$i=0$ \TO $M$}
   \STATE $\log{\mathcal{O}} = \log{\mathcal{O}} - \log{\left(\sqrt{\texttt{S}[i]}\right)}$
 \ENDFOR
 \STATE $\log{\mathcal{O}} = \log{\mathcal{O}} + N\log{\sigma} - (Z - Y^2/4X)/2\sigma^2$
 \IF{integrating the final amplitude from $[0, \infty]$}
  \STATE $\log{\mathcal{O}} = \log{\mathcal{O}} + M\log{\left(\sqrt{2\pi}\right)} +
\log{\left(\sqrt{\pi/2}\right)}$
  \STATE $\log{\mathcal{O}} = \log{\mathcal{O}} + \log{ \left( \erfc{
\left(Y/\left(2\sqrt{2\sigma^2X}\right)\right) } \right) }$
 \ELSE[integrating the final amplitude from $-\infty$ to $\infty$]
  \STATE $\log{\mathcal{O}} = \log{\mathcal{O}} + N\log{\left(\sqrt{2\pi}\right)}$
 \ENDIF
 \STATE add $\log{({\rm amplitude~priors})}$ to odds ratio
\end{algorithmic}

\end{algorithm}

\subsection{Marginalization without the final amplitude}\label{app:margpe}

In the above marginalizations all the component amplitudes have been marginalized. However, if, for
example, performing parameter estimation one might be interested in the actual posterior probability
distribution of one of those amplitudes, or there might be some subset of components that do not
have a factorizable amplitude, which therefore cannot be marginalized over with this algorithm.
These require a slight modification of Algorithm~\ref{alg:logO} as given in
Algorithm~\ref{alg:notfinal}, where essentially the final amplitude integral is not performed. In
our case when performing parameter estimation we would be interested in the actual posterior
probability distribution of the flare amplitude, but would still want the background polynomial
coefficients marginalized over.
\begin{algorithm}
\caption{Calculate $\log{\mathcal{O}}$ marginalized over model component amplitudes except for a
final model component, where the sizes of \texttt{C} and \texttt{S} are the same as for
Algorithm~\ref{alg:logO} (again this assumes \texttt{C}-style array indexing and that $N$ refers to
the total number of model components including the one being left out of the marginalisation).}
\label{alg:notfinal}
\begin{algorithmic}[1]
 \STATE $Z = 0$
 \STATE $Y = 0$
 \STATE $M = N-1$
 \STATE $P = N-2$
 \STATE $\log{\mathcal{O}} = 0$
 \FOR{$i=0$ \TO $P$}
   \FOR{$j=i$ \TO $M$}
     \FOR{$k=i$ \TO $M$}
       \IF{$j \neq i+1$ \AND $j \neq P$ \AND $k \neq i+1$ \AND $k \neq P$}
         \STATE $Z = Z - \texttt{C}[i][j]\times\texttt{C}[i][k]/4\texttt{S}[i]$
       \ENDIF
       \IF{$j$ is equal to $i$}
         \IF{$k > j$ \AND $k < M$}
           \STATE $\texttt{S}[k] = \texttt{S}[k] - \texttt{C}[j][k]^2/4\texttt{S}[i]$
         \ENDIF
       \ELSE
         \IF{$k$ is equal to $i$ \AND $j < M$}
           \STATE $\texttt{C}[j][j] = \texttt{C}[j][j] - \texttt{C}[i][j]\times
\texttt{C}[i][k]/2\texttt{S}[i]$
         \ELSIF{$k > j$}
           \STATE $\texttt{C}[j][k] = \texttt{C}[j][k] - \texttt{C}[i][j]\times
\texttt{C}[i][k]/2\texttt{S}[i]$
         \ENDIF
       \ENDIF
     \ENDFOR
   \ENDFOR
 \ENDFOR
 \STATE $X = \texttt{S}[P]$
 \FOR{$i=P$ \TO $M$}
   \STATE $Y = Y + \texttt{C}[P][i]$
 \ENDFOR
 \STATE $Z = Z + \texttt{S}[M] + \texttt{C}[M][M]$
 \FOR{$i=0$ \TO $P$}
   \STATE $\log{\mathcal{O}} = \log{\mathcal{O}} - \log{\left(\sqrt{\texttt{S}[i]}\right)}$
 \ENDFOR
 \STATE $\log{\mathcal{O}} = \log{\mathcal{O}} + M\log{\left(\sqrt{2\pi\sigma^2}\right)} - (Z -
Y^2/4X)/2\sigma^2$
 \STATE add $\log{({\rm amplitude~priors})}$ to odds ratio
\end{algorithmic}

\end{algorithm}

\section{An alternative method for noise estimation}\label{app:noiseest}

Here we describe an alternative to the noise estimation method discussed in
Section~\ref{sec:noise}. Flares have most of their spectral power at low frequencies, and also the
background variations are at low frequencies (compared to the sample time), so assuming that the
underlying noise is white and Gaussian the high-frequency end of the spectrum can be used to
estimate the noise standard deviation. For spectra of light curves with very large amplitude
variations a large amount of the power from the variations can leak into the high-frequency part of
the spectrum and lead to an overestimation for the noise, so the data must first be detrended
using e.g.\ the Savitzky-Golay filtering algorithm. We then take the mean of the final half of the
one-sided power spectrum of the filtered data $\bar{S}_n$ (although different fractions of the
spectrum could be used) and get the time series standard deviation via $\sigma =
\sqrt{\bar{S}_n/(2\Delta t)}$, where $\Delta{}t$ is the time-step between points (in seconds).

For flare signals with \snr of a few 10s their power can start leaking into the part of the
spectrum we use for noise estimation. This would mean that for data containing loud flares there
would be a bias against finding quieter flares due to noise overestimation.

\end{document}